\documentclass[11pt,a4paper]{article}
\usepackage{epsfig,axodraw}
\usepackage{array,cite,amsmath,amssymb}
\usepackage[flushleft]{threeparttable}
\usepackage{subcaption}
\usepackage{kantlipsum}
\usepackage{float}
\usepackage[usenames,dvipsnames]{xcolor}
\usepackage[breakable, theorems, skins]{tcolorbox}
\tcbset{enhanced}
\usepackage{booktabs}
\usepackage{caption}
\usepackage{threeparttable}
\usepackage{multirow}
\usepackage{listings}
\definecolor{DarkMagenta}{rgb}{0.55,0.0,0.55}
\newcommand{\newc}{\newcommand}
\newc{\gev}{\,GeV}
\newcolumntype{M}[1]{>{\centering\arraybackslash}m{#1}}
\newcolumntype{N}{@{}m{0pt}@{}}
\newc{\mev}{\,MeV}
\newc{\ra}{\rightarrow}
\newc{\rpv}{$\mathrm{\not\!R_p}$}
\newc{\rp}{$\mathrm{R_p}$}
\newc{\real}{\mathcal{R}e}
\newc{\alsm}{{\displaystyle \sum_{\alpha=1,2}}}
\newc{\besm}{{\displaystyle \sum_{\beta=1,2}}}
\newc{\al}{\alpha}
\newc{\sgn}{\mr{sgn}\,}
\newc{\be}{\beta}
\newc{\ga}{\gamma}
\newc{\de}{\delta}
\newc{\sla}{\!\!\!\!\!\not\:\:\!}
\newc{\slab}{\!\!\!\!\!\not\,\,\,}
\newc{\slac}{\!\!\!\!\!\!\!\not\,\,\,\,}
\newc{\met}{$\not\!\!E_T$}
\newc{\cw}{\cos\theta_W}
\newc{\sw}{\sin\theta_W}
\newc{\ssw}{\sin^2\theta_W}
\newc{\ccw}{\cos^2\theta_W}
\newc{\cbe}{\cos\beta}
\newc{\sbe}{\sin\beta}
\newc{\ort}{\frac1{\sqrt{2}}}
\newc{\sh}{\hat{s}}
\newc{\uh}{\hat{u}}
\newc{\tha}{\hat{t}}
\newc{\sa}{\sin\al}
\newc{\ca}{\cos\al}
\newc{\mz}{M_{\mr{Z}}}
\newc{\mw}{M_{\mr{W}}}
\newc{\bv}{$\mathrm{\not\!B}$}
\newc{\lv}{$\mathrm{\not\!L}$}
\newc{\beq}{\begin{equation}}
\newc{\eeq}{\end{equation}}
\newc{\ie}{{\it i.e.\/}\ }
\newc{\lam}{\lambda}
\newc{\cht}{\tilde{\chi}}
\newc{\glt}{\tilde{g}}
\newc{\upt}{\tilde{u}}
\newc{\qkt}{\tilde{q}}
\newc{\elt}{\tilde{\ell}}
\newc{\hgt}{\tilde{H}}
\newc{\nut}{\tilde{\nu}}
\newc{\dnt}{\tilde{d}}
\newc{\ftl}{\mr{\tilde{f}}}
\newc{\psb}{\bar{\psi}}
\newc{\rtt}{2^{1/2}}
\newc{\mut}{\tilde{\mu}}
\newc{\mr}{\mathrm}
\newc{\bath}{\bar{\theta}}
\newc{\tht}{\theta}
\newc{\JC}{{\bf J}}
\newc{\lra}{\longrightarrow}
\newc{\eg}{{\it e.g.\  }}
\newc{\barr}{\begin{eqnarray}}
\newc{\earr}{\end{eqnarray}}
\newc{\me}{\mathcal{M}}
\newc{\dbm}{\partial_\mu}
\newc{\dbmu}{\stackrel{\leftrightarrow\  }{\partial^\mu}}
\newc{\sgm}{\sigma_\mu}
\newc{\captionB}[2]{\caption[{#1}]{{\small {#2}}}}
\newc{\ahref}[2]{#2}
\pdfoutput=1 

\usepackage{array,cite,amsmath,color}
\usepackage{xspace}
\usepackage{bbm,subcaption}
\usepackage{slashed}

\usepackage{jheppub} 
\title{Gauge-Invariant Longitudinal Modes in the Herwig 7 Electroweak Parton Shower}

\author{M.R.~Masouminia,}
\author{P.~Richardson}

\affiliation{Institute for Particle Physics Phenomenology, Durham University, Durham, UK}

\emailAdd{mohammad.r.masouminia@durham.ac.uk}
\emailAdd{peter.richardson@durham.ac.uk}

\abstract{ \small
Longitudinal electroweak gauge bosons are the most technically delicate ingredient of electroweak parton showers: in the broken Standard Model, the gauge component of a longitudinal polarisation does not cancel diagram by diagram, but is related by Ward identities to amplitudes with an insertion of the associated would-be Goldstone field. Building on the default \textsf{Herwig~7} treatment based on a subtraction-defined longitudinal current, we construct a gauge-invariant scheme in which the subtraction remainder is retained as a well-defined contribution and completed by a Ward-identity-fixed Goldstone-matching term. We derive helicity-resolved building blocks and compact quasi-collinear splitting kernels for $q\to q'V$ and $V\to V'V''$ branchings, and implement the scheme in \textsf{Herwig~7} as a switchable alternative to the default basis. The completion leaves the transverse sector unchanged and modifies only longitudinal entries through controlled symmetry-breaking terms, including Yukawa-sensitive contributions in massive-fermion channels. In shower-level studies, we find that the default and gauge-invariant prescriptions coincide at high evolution scales, while differing at lower scales precisely in channels where symmetry breaking is active. Exclusive single-emission observables can moreover display non-monotonic scheme dependence once Sudakov suppression and kinematic constraints are accounted for. A first multi-emission LHC-like study confirms numerical stability and yields controlled, interpretable shifts in observables that probe propagating electroweak shower currents, whereas quantities dominated by promptly contracted, near on-shell vector-boson matrix elements remain largely unchanged.
}

\begin{document}

\noindent{\hfill \small IPPP/25/91\\[0.1in]}
\maketitle
\flushbottom

\section{Introduction}
\label{sec:intro}

% importance of EW parton shower, difficulties in longitudinal modes and state-of-the-art treatment of them
Electroweak (EW) parton showers have become an essential component of precision and discovery phenomenology at present and future hadron colliders, because in the multi-TeV regime the exclusive structure of events is increasingly shaped by real and virtual EW radiation and by the associated Sudakov-enhanced logarithms, schematically $\alpha_{\mathrm{w}}\ln^{2}(Q^{2}/M_{W}^{2})$ and $\alpha_{\mathrm{w}}\ln(Q^{2}/M_{W}^{2})$, which can compete numerically with QCD effects in energetic tails and are not reliably captured by fixed-order calculations alone once widely separated scales and multiple emissions are relevant \cite{Ciafaloni:2000df,Denner:2000jv, Kuhn:1999nn}. Implementing this resummation in a fully differential Monte Carlo framework, however, is more subtle than in QCD: the broken gauge theory introduces physical masses, chiral couplings, and mixing, and in particular renders the treatment of longitudinal massive vector bosons delicate. The conventional polarisation vector $\varepsilon_{L}^{\mu}(p)$ contains a formally large $p^{\mu}/M_{V}$ component whose contributions are gauge artefacts that cancel only after enforcing the appropriate Ward identities, with the cancellation controlled, at leading power in $M_{V}^{2}/Q^{2}$, by the Goldstone-boson equivalence theorem and by Yukawa interactions for massive fermions \cite{Cornwall:1974km,Chanowitz:1985hj, Denner:2001gw}. In a shower context, where one factorises a hard process from quasi-collinear branchings and interprets the latter probabilistically, any ad hoc subtraction or redefinition of $\varepsilon_{L}^{\mu}$ that removes the $p^{\mu}/M_{V}$ term must therefore be accompanied by a correspondingly consistent inclusion of the would-be Goldstone contribution; otherwise one risks violating gauge invariance at the level of the splitting operator and, practically, suppressing or mis-modelling such configurations precisely in the ultracollinear region (with transverse momenta of order $M_V$, where mass effects are parametrically important) \cite{Masouminia:2021kne, Dittmaier:2025htf}. The state-of-the-art consequently combines helicity-resolved (or spin-density-matrix) evolution with quasi-collinear EW splitting kernels in the broken theory, together with prescriptions, either in physical gauges or via explicit Goldstone-matching, that maintain gauge-invariant (GI) longitudinal currents at the level of the splitting operator while retaining the correct on-shell/virtual interpolation for $W/Z/H$ emissions and preserving spin correlations to the accuracy claimed by the shower \cite{Masouminia:2021kne, Christiansen:2014kba, Krauss:2014yaa, Bauer:2017isx}.

% summary of EW PS in Herwig 7 [2108.10817] and the use of SL's picture
In \textsf{Herwig~7}~\cite{Bahr:2008pv, Bellm:2015jjp, Bellm:2017bvx, Bellm:2019zci, Bewick:2023tfi, Bellm:2025pcw}, the EW parton shower was formulated as an extension of the angular-ordered, spin-correlated shower to the spontaneously broken $\mathrm{SU}(2)_L\times \mathrm{U}(1)_Y$ theory, with quasi-collinear splitting operators for fermion, vector-boson and scalar lines embedded into the evolution via spin-density matrices in order to retain leading spin correlations within a Markovian branching framework \cite{Masouminia:2021kne,Darvishi:2021het,Bellm:2025pcw}. The longitudinal sector was handled by adopting a subtraction-longitudinal (SL) picture (also known as Dawson's approach~\cite{Dawson:1984gx}): longitudinal kernels were constructed using a manifestly finite definition of the longitudinal state which avoids the explicit appearance of the $p^\mu/M_V$ component of $\varepsilon_L^\mu(p)$ at the level of the branching operator, thereby providing numerically stable helicity-dependent splitting functions suitable for shower evolution \cite{Masouminia:2021kne}. This prescription is well-motivated in Equivalent Vector Boson Approximation (EWA)-like kinematics, and fixed-order comparisons in that regime provided non-trivial validation of the resulting resummation pattern \cite{Masouminia:2021kne}; nevertheless, recent work has emphasised that, outside this restricted limit, the structure of longitudinal contributions in the broken theory is constrained by Ward identities which relate the would-be $p^\mu$ contraction to amplitudes involving the corresponding Goldstone mode (and, for massive fermions, Yukawa interactions), raising the question of whether a universal application of the SL construction across all EW branchings can miss parts of the ultracollinear longitudinal dynamics \cite{Dittmaier:2025htf}. The resulting discussion is therefore not about the practicality of the SL implementation, but about its domain of formal validity and the extent to which a GI completion is required to control longitudinal rates and polarisation systematics in general shower applications.

% SL's picture vs gauge-invariant picture
Conceptually, it is useful to distinguish between the SL picture and a fully GI treatment of longitudinal modes. In the SL picture, one replaces the conventional polarisation vector by a subtracted object, $\varepsilon_L^\mu(p)\to \varepsilon_{L*}^\mu(p)$, obtained by removing the component proportional to $p^\mu/M_V$, so that the splitting currents entering the quasi-collinear kernels are manifestly finite and directly amenable to a probabilistic shower interpretation \cite{Masouminia:2021kne}; this construction is particularly natural in EWA-like kinematics, where the subtracted $p^\mu/M_V$ contribution is power-suppressed or eliminated once the appropriate Ward identities are enforced on the hard amplitude \cite{Masouminia:2021kne}. In the spontaneously broken theory, however, the relevant Ward identities do not, in general, imply that contractions with $p^\mu$ vanish: instead, they relate them to amplitudes with an insertion of the associated would-be Goldstone field (with Yukawa-suppressed contributions for massive fermions), so that the physical limit is governed by a specific vector-Goldstone combination \cite{Cornwall:1974km,Chanowitz:1985hj,Denner:2001gw}. A GI longitudinal scheme therefore cannot be realised by subtraction alone; rather, it promotes the subtraction remainder to one component of the physical longitudinal current and supplements it by the Ward-identity-fixed Goldstone-matching term (for our conventions), yielding a longitudinal splitting operator which is gauge invariant by construction and applicable to both on-shell and off-shell shower kinematics across the full set of EW branching channels \cite{Dittmaier:2025htf}.

% what we are doing in this paper
In this paper, we formulate and implement such a GI completion of the longitudinal sector of the \textsf{Herwig~7} electroweak parton shower. We derive the corresponding helicity-resolved building blocks and compact quasi-collinear splitting kernels for both fermion and vector-boson branchings, and implement the scheme in \textsf{Herwig~7} in a manner which preserves backwards compatibility and allows a transparent comparison between the SL and GI pictures within an otherwise identical shower setup. We validate the construction by analytic checks of the relevant Ward identities and by numerical comparisons at the level of splitting functions and controlled shower evolutions, isolating the regions of phase space in which the GI completion affects longitudinal rates and kinematics; the formal construction is presented in Sec.~\ref{sec:formalism}.

% outlook of the paper (updated to match section plan)
The remainder of this paper is organised as follows. In Sec.~\ref{sec:kinematics} we summarise the \textsf{Herwig~7} conventions for shower kinematics and evolution variables relevant for the EW shower, including the dot-product preserving reconstruction used throughout our validation studies. In Sec.~\ref{sec:formalism} we develop the GI treatment of longitudinal polarisations in the broken theory, formulating a longitudinal splitting operator consistent with the relevant Ward identities and the associated Goldstone-matching, and specifying its relation to the SL picture used previously. Our results are presented in Sec.~\ref{sec:res}: we first compare the SL and GI schemes at the level of helicity building blocks and analytic quasi-collinear splitting functions, and then study their impact within controlled shower evolutions, ranging from single-resummed tests to a first multi-emission LHC-like check, with an emphasis on longitudinally sensitive observables and polarisation systematics. We conclude in Sec.~\ref{sec:conc}. Details of the \textsf{Herwig} user interface and steering of the longitudinal scheme are collected in App.~\ref{sec:ui-long}.

\section{Herwig~7 Conventions for Parton Shower Kinematics}
\label{sec:kinematics}

We follow the kinematic conventions of the angular-ordered shower and, unless stated otherwise, adopt the \emph{dot-product preserving} reconstruction scheme, in which the evolution variable is identified with the (mass-corrected) daughter dot product and is kept fixed under subsequent evolution of the daughters \cite{Bewick:2019rbu,Bewick:2023tfi}.  This choice provides a recoil prescription which avoids excessive growth of parent virtuality under multiple emissions while maintaining the logarithmic accuracy of the angular-ordered formulation.  We use the metric signature $(+,-,-,-)$.

We work in a frame in which the pre-branching parent is aligned with $+\hat z$ and introduce a lightlike reference vector $n^\mu$,
\begin{equation}
p^\mu=\big(E_p;\,0,0,p\big),\qquad E_p=\sqrt{p^2+m_0^2},\qquad
n^\mu=(1,0,0,-1),\qquad n^2=0,\qquad p\!\cdot\!n>0,
\end{equation}
where $p\equiv|\vec p|$ denotes the parent three-momentum magnitude and $m_0$ its pole mass.  In a $0\to 1\,2$ branching, momenta are decomposed as
\begin{align}
q_1^\mu &= z\,p^\mu+\beta_1\,n^\mu+q_\perp^\mu, \nonumber\\
q_2^\mu &= (1-z)\,p^\mu+\beta_2\,n^\mu-q_\perp^\mu, \nonumber\\
q_\perp^\mu &= (0;\,p_\perp\cos\varphi,\,p_\perp\sin\varphi,\,0),
\label{eq:q12}
\end{align}
with $0<z<1$ and $p\!\cdot\!q_\perp=n\!\cdot\!q_\perp=0$.  Imposing on-shell conditions $q_1^2=m_1^2$ and $q_2^2=m_2^2$ yields
\begin{equation}
\beta_1=\frac{p_\perp^2+m_1^2-z^2 m_0^2}{2z\,p\!\cdot\!n},\qquad
\beta_2=\frac{p_\perp^2+m_2^2-(1-z)^2 m_0^2}{2(1-z)\,p\!\cdot\!n},
\label{eq:betas}
\end{equation}
and the off-shell parent entering the splitting is
\begin{equation}
q_0^\mu=p^\mu+\beta_0\,n^\mu,\qquad \beta_0=\beta_1+\beta_2,\qquad
q_0^2=\frac{p_\perp^2}{z(1-z)}+\frac{m_1^2}{z}+\frac{m_2^2}{1-z}\,.
\label{eq:hw7-parent}
\end{equation}
To organise leading powers in the quasi-collinear limit, we introduce a bookkeeping parameter $\lambda$ and rescale
\begin{equation}
m_i\to\lambda m_i,\qquad p_\perp\to\lambda p_\perp,
\end{equation}
keeping $p$ and $z$ fixed; expansions are taken consistently in $\lambda$ to the minimal order required for cancellations.

The shower generates a pair $(z,\tilde q^2)$ and reconstructs the branching kinematics.  For on-shell daughters, there are several equivalent representations of the evolution variable, which coincide up to mass terms; in particular, one may write
\begin{equation}
\tilde q^2
=\frac{2\,q_1\!\cdot q_2+m_1^2+m_2^2-m_0^2}{z(1-z)}
\equiv \frac{q_0^2-m_0^2}{z(1-z)},
\label{eq:hw7-qtilde}
\end{equation}
where the first form is the natural starting point for the dot-product preserving scheme.  When the daughters subsequently acquire virtualities, the three forms in \eqref{eq:hw7-qtilde} no longer coincide, and a reconstruction prescription is required; in the dot-product preserving scheme, one keeps the dot-product form fixed and reconstructs the transverse momentum accordingly.  Allowing for off-shell daughters, this gives
\begin{equation}
p_\perp^2
= z^2(1-z)^2\,\tilde q^2 \;-\; (1-z)^2\,q_1^2 \;-\; z^2\,q_2^2
\;+\; z(1-z)\big(m_0^2-m_1^2-m_2^2\big),
\label{eq:hw7-pt2}
\end{equation}
which for $q_i^2=m_i^2$ becomes $p_\perp^2=z^2(1-z)^2\tilde q^2+z(1-z)m_0^2-(1-z)m_1^2-zm_2^2$, and reduces to $p_\perp^2=z^2(1-z)^2\tilde q^2$ in the massless limit. A sufficient (though not necessary) physicality condition ensuring that a real solution for $p_\perp$ exists after subsequent emissions is
\begin{equation}
\tilde q^2 \;>\; 2\,\max\!\left(\frac{m_1^2}{z^2},\frac{m_2^2}{(1-z)^2}\right),
\qquad\Rightarrow\qquad p_\perp^2\ge 0\ \text{at reconstruction}
\label{eq:hw7-bound}
\end{equation}
In practice, the shower may still enter regions where the reconstructed $p_\perp^2$ would be negative, in which case the standard \textsf{Herwig} treatment is applied (veto/replacement depending on scheme and settings), but \eqref{eq:hw7-bound} is adequate for analytic power counting and for identifying the parametrically allowed phase space.

For backward evolution $\tilde\imath j\to i+j$ with $i$ space-like and $j$ time-like, we employ the analogous dot-product preserving definition of the ordering variable,
\begin{equation}
\tilde q^2 \equiv \frac{2\,p_j\!\cdot p_{\tilde\imath j}-m_j^2}{1-z},
\label{eq:hw7-isr-qtilde}
\end{equation}
together with the corresponding transverse-momentum reconstruction
\begin{equation}
p_\perp^2=(1-z)^2\tilde q^2-p_j^2-(1-z)^2 p_{\tilde\imath j}^2+(1-z)m_j^2,
\label{eq:hw7-isr-pt2}
\end{equation}
where $z$ is the lightcone momentum fraction taken from the incoming leg \cite{Bewick:2021nhc}.  In the massless limit, this ordering variable can also be written in the equivalent forms
\begin{equation}
\tilde q^2=\frac{p_\perp^2}{(1-z)^2}=\frac{-q_1^2}{1-z}=\frac{2\,q_0\!\cdot q_2}{1-z},
\end{equation}
illustrating the close correspondence between the dot-product definition and the backwards-evolution virtuality assignment.

Finally, for reference, our symbols are: $p$ (scalar) is the parent three-momentum magnitude in the aligned frame; $p^\mu$ and $n^\mu$ are reference four-vectors defining the Sudakov decomposition; $q_0,q_1,q_2$ are the parent and daughter four-momenta; $m_0,m_1,m_2$ are pole masses; $z\in(0,1)$ is the $n$-lightcone fraction of child~1; $p_\perp\ge0$ and $\varphi$ are the modulus and azimuth of the transverse recoil; and $\tilde q$ is the angular-ordering evolution variable defined in \eqref{eq:hw7-qtilde}.

\section{Quasi-Collinear Helicity-Dependent Splitting Functions}
\label{sec:formalism}

Our starting point is the quasi-collinear factorisation of a generic shower branching $0\to 1\,2$, expressed in terms of helicity-resolved splitting amplitudes evaluated with the kinematics and evolution variable introduced in Sec.~\ref{sec:kinematics}.  Throughout we work in the aligned frame where the pre-branching parent momentum is along $+\hat z$, and organise the quasi-collinear expansion by the rescaling $m_i\to \lambda m_i$ and $p_\perp\to \lambda p_\perp$ at fixed $p$ and $z$, retaining the minimal order in $\lambda$ required for finiteness after all gauge- and symmetry-breaking cancellations are enforced.  In practice this means that we compute the helicity amplitudes with explicit spinors and polarisation vectors expanded consistently in $\lambda$, using the same conventions as the original \textsf{Herwig} EW shower construction \cite{Masouminia:2021kne} so that the connection to the implemented kernels is manifest, while the dot-product preserving reconstruction of the branching kinematics ensures that the invariants entering the helicity algebra coincide with those realised by the multi-emission momentum map \cite{Bewick:2019rbu, Bewick:2021nhc, Bewick:2023tfi, Bellm:2025pcw}.

For definiteness, consider first $q\to q'V$ with $V=W^\pm,Z^0$ and the emitted boson labelled as particle~$2$.  A convenient explicit representation for the transverse polarisation vectors of the emitted vector boson in the aligned frame is \cite{Masouminia:2021kne}
\begin{equation}
\epsilon^\mu_{\lambda_2=\pm 1}(q_2)=
\left[
0;\;
-\frac{\lambda_2}{\sqrt{2}}
\left(
1-\frac{p_\perp^2\lambda^2 e^{i\lambda_2\phi}\cos\phi}{2p^2(1-z)^2}
\right),\;
-\frac{i}{\sqrt{2}}+\frac{\lambda_2 p_\perp^2\lambda^2 e^{i\phi}\sin\phi}{2\sqrt{2}p^2(1-z)^2},\;
-\frac{\lambda_2 p_\perp\lambda e^{i\lambda_2\phi}}{\sqrt{2}(1-z)p}
\right].
\label{eq:prelude-epsT}
\end{equation}
For the fermionic line, we employ explicit helicity spinors for the incoming (parent) and outgoing (child) quarks, expanded in $\lambda$ in the same frame.  With $p$ the parent 3-momentum magnitude and $m_0$ the parent pole mass, the incoming spinors may be taken as \cite{Masouminia:2021kne}
\begin{equation}
u_{\frac{1}{2}}(p)=
\left(
\begin{array}{c}
\frac{m_0}{\sqrt{2p}}\lambda\\[2pt]
0\\[2pt]
\sqrt{2p}\left(1+\frac{m_0^2\lambda^2}{8p^2}\right)\\[2pt]
0
\end{array}
\right),
\qquad
u_{-\frac{1}{2}}(p)=
\left(
\begin{array}{c}
0\\[2pt]
\sqrt{2p}\left(1+\frac{m_0^2\lambda^2}{8p^2}\right)\\[2pt]
0\\[2pt]
\frac{m_0}{\sqrt{2p}}\lambda
\end{array}
\right),
\label{eq:prelude-u-in}
\end{equation}
and for the outgoing fermion of momentum $q_1$ (lightcone fraction $z$) one may use \cite{Masouminia:2021kne}
\begin{align}
\bar u_{\frac{1}{2}}(q_1)
&=
\left[
\sqrt{2zp}\left(1+\frac{m_0^2\lambda^2}{8p^2}\right),\;
\frac{e^{-i\phi}p_\perp\lambda}{\sqrt{2zp}},\;
\frac{m_1}{\sqrt{2zp}}\lambda,\;
\frac{e^{-i\phi}p_\perp m_1\lambda^2}{(2zp)^{3/2}}
\right],
\nonumber\\
\bar u_{-\frac{1}{2}}(q_1)
&=
\left[
-\frac{e^{i\phi}p_\perp m_1\lambda^2}{(2zp)^{3/2}},\;
\frac{m_1}{\sqrt{2zp}}\lambda,\;
-\frac{e^{i\phi}p_\perp\lambda}{\sqrt{2zp}},\;
\sqrt{2zp}\left(1+\frac{m_0^2\lambda^2}{8zp}\right)
\right],
\label{eq:prelude-ubar-out}
\end{align}
where $m_1$ is the pole mass of the outgoing fermion (typically $m_1=m_0$ for flavour-diagonal branchings), and $\phi$ is the azimuth of the transverse recoil defined in Sec.~\ref{sec:kinematics}.  Analogous expressions apply to antifermions and to backward evolution, with the same lightcone parametrisation and evolution variable as used by the shower \cite{Masouminia:2021kne}.

For $V\to V'V''$ branchings, we similarly require explicit polarisation vectors for a potentially massive parent vector.  In the same aligned frame, a standard choice for the parent transverse and longitudinal polarisation vectors is \cite{Masouminia:2021kne}
\begin{equation}
\epsilon^\mu_{\lambda_0=\pm 1}(p)=\left[0,\;-\frac{\lambda_0}{\sqrt{2}},\;-\frac{i}{\sqrt{2}},\;0\right],
\qquad
\epsilon^\mu_{0}(p)=\left[\frac{p}{\lambda m_0},\;0,\;0,\;\frac{\sqrt{\lambda^2 m_0^2+p^2}}{\lambda m_0}\right],
\label{eq:prelude-eps-parent}
\end{equation}
with $m_0$ the parent pole mass.  The longitudinal vector $\epsilon^\mu_0(p)$ contains the familiar $p^\mu/m_0$ behaviour and is therefore the locus of the longitudinal-mode subtleties in the broken theory.  The \textsf{Herwig~7} EW shower, as originally formulated, instead employs a manifestly finite SL remainder, obtained by subtracting the would-be gauge piece and which in this frame can be written as \cite{Masouminia:2021kne}
\begin{equation}
\epsilon^\mu_{0_*}(p)=
\left[
-\frac{\lambda m_0}{p+\sqrt{\lambda^2m_0^2+p^2}},\;
0,\;0,\;
\frac{\lambda m_0}{p+\sqrt{\lambda^2m_0^2+p^2}}
\right].
\label{eq:prelude-eps-SL}
\end{equation}
The child-vector polarisation vectors are taken consistently with the same quasi-collinear expansion and with the momentum mapping of Sec.~\ref{sec:kinematics}, so that the helicity algebra is performed on the same invariants that define $\tilde q$ and the reconstructed $p_\perp$ used in the shower \cite{Masouminia:2021kne}.

Given these building blocks, the splitting kernels are obtained from reduced helicity amplitudes $H_{\lambda_0,\lambda_1,\lambda_2}(z,\tilde q)$ defined by stripping the universal collinear normalisation from the corresponding $1\to 2$ matrix element in the quasi-collinear limit, in direct analogy with the construction in Ref.~\cite{Masouminia:2021kne}.  The (spin-dependent) branching probability can then be written in the generic form
\begin{equation}
\mathrm{d}\mathcal{P}_{0\to 1\,2}
=
\frac{\alpha_{\mathrm{int}}(\tilde q^2)}{2\pi}\,
\frac{\mathrm{d}\tilde q^2}{\tilde q^2}\,
\mathrm{d}z\;
F_{0\to 1\,2}(z,\tilde q^2;\rho_0),
\label{eq:prelude-dP}
\end{equation}
with $\rho_0$ the spin-density matrix of the emitter (or parent), and where the spin-correlated kernel entering the Markovian evolution is constructed from the helicity amplitudes as
\begin{equation}
F_{0\to 1\,2}(z,\tilde q^2;\rho_0)
=
\sum_{\lambda_0,\lambda_0',\lambda_1,\lambda_2}
\rho_0^{\lambda_0\lambda_0'}\,
H_{\lambda_0,\lambda_1,\lambda_2}(z,\tilde q)\,
H^{*}_{\lambda_0',\lambda_1,\lambda_2}(z,\tilde q),
\label{eq:prelude-kernel}
\end{equation}
which reduces to the familiar unpolarised splitting function upon averaging $\rho_0$ over the parent helicities.  The decomposition of $F_{0\to 1\,2}$ into transverse-transverse, longitudinal-longitudinal, and mixed transverse-longitudinal components is thus entirely controlled by the choice of longitudinal basis used to define $H$, and this is precisely where the distinction between the SL prescription and a GI longitudinal construction enters.  In the SL picture, one effectively defines the longitudinal contribution by the replacement $\epsilon_0^\mu\to \epsilon_{0_*}^\mu$ at the level of the splitting current, thereby yielding a finite longitudinal kernel that is straightforwardly embedded into the shower \cite{Masouminia:2021kne}.  In a fully GI treatment, by contrast, the longitudinal splitting current is defined such that it obeys the broken-theory Ward identities, which relate contractions of the vector current with the boson momentum to amplitudes with insertions of the corresponding would-be Goldstone field (and Yukawa terms for massive fermions), implying that the physically meaningful longitudinal object is a specific vector-Goldstone combination rather than the subtraction remainder alone \cite{Borel:2012by,Cuomo:2019siu,Dittmaier:2025htf}.  The following two subsections make this statement concrete for $q\to q'V$ and $V\to V'V''$ respectively, and provide the explicit SL and GI definitions of the reduced helicity amplitudes $H$ from which the shower kernels are constructed. Splittings involving a physical Higgs, $q\to q'S$ and $V\to V'S$ ($S$ being any possible scalar boson), do not require a SL/Goldstone completion of a longitudinal current in the sense discussed above, and their phenomenology is controlled by the corresponding Yukawa- and mass-suppressed structures.

\subsection{$q\to q'V$ splittings}
\label{sec:qqV}

We consider the generic electroweak branching
\begin{equation}
q(p_0,\lambda_0)\;\longrightarrow\; q'(p_1,\lambda_1)\;+\;V(p_2,\lambda_2)\,,
\end{equation}
with $V=W^\pm,Z^0$ a massive gauge boson and helicities $\lambda_0,\lambda_1=\pm\frac{1}{2}$, $\lambda_2=\pm,0$.  The interaction is parametrised by the chirality-projected vector current and the associated Goldstone Yukawa current,
\begin{align}
\mathcal{L}_{qqV} &\supset \bar q'\gamma^\mu\left(g_L P_L+g_R P_R\right) q\,V_\mu + \mathrm{h.c.}\,, \label{eq:qqV-Lvec}\\
\mathcal{L}_{qqG} &\supset \bar q'\left(y_L P_L+y_R P_R\right) q\,G_V + \mathrm{h.c.}\,, \label{eq:qqV-Lgol}
\end{align}
where $G_V$ denotes the would-be Goldstone mode associated with $V$, and $P_{L,R}=(1\mp\gamma_5)/2$.  In the Standard Model, one has, for example, for charged-current emission $q_u\to q_d W^+$,
\begin{equation}
g_L=\frac{g}{\sqrt{2}}V_{ud},\qquad g_R=0,\qquad
y_L=\frac{g\,m_u}{\sqrt{2}M_W}V_{ud},\qquad
y_R=-\frac{g\,m_d}{\sqrt{2}M_W}V_{ud},
\label{eq:qqV-SM}
\end{equation}
and analogous relations for $Z^0/G^0$ with the appropriate chiral charges. Throughout, we use the \textsf{Herwig~7} angular-ordered kinematics of Sec.~\ref{sec:kinematics} with
\begin{equation}
t=\tilde q^{\,2}\,z(1-z),
\qquad
p_\perp^2=t\,z(1-z) + z(1-z)m_0^2-(1-z)m_1^2-zm_2^2.
\end{equation}
We follow the quasi-collinear construction and spin-density-matrix embedding of Ref.~\cite{Masouminia:2021kne}.  For a generic chiral electroweak vertex,
\begin{equation}
\mathcal{O}[V] \;\propto\; \bar q'(q_1)\,\gamma^\mu\big(g_L P_L + g_R P_R\big)\,q(q_0)\,V_\mu(q_2)\,,
\end{equation}
The helicity-dependent branching operator is built from reduced amplitudes
\begin{equation}
\widehat H_{\lambda_0\lambda_1\lambda_2}
\;=\;
\frac{1}{\mathcal{D}}\,
\bar u_{\lambda_1}(q_1)\,\gamma^\mu\big(g_L P_L + g_R P_R\big)\,u_{\lambda_0}(q_0)\,
\varepsilon^{(\mathrm{X})*}_{\mu,\lambda_2}(q_2)\,,
\qquad
\mathcal{D}\equiv \sqrt{2\,(q_0^2-m_0^2)}\,,
\end{equation}
with $\mathrm{X}=\mathrm{SL},\mathrm{GI}$ denoting the longitudinal prescription and with $\varepsilon_{\mu,\lambda_2}$ evaluated using the kinematics and phase conventions of Sec.~\ref{sec:kinematics}.  In \textsf{Herwig~7}, the corresponding splitting kernel is obtained by contracting $\widehat H$ with the fermion spin-density matrix $\rho$ of the emitting line and (optionally) retaining a spin-density matrix for the emitted vector for subsequent evolution/decay,
\begin{equation}
F_{q\to q'V}^{(\mathrm{X})}(z,\tilde q^2)
\;\propto\;
\sum_{\lambda_0,\lambda_0',\lambda_1,\lambda_2,\lambda_2'}
\rho_{\lambda_0\lambda_0'}\,
\widehat H^{(\mathrm{X})}_{\lambda_0\lambda_1\lambda_2}\,
\widehat H^{(\mathrm{X})*}_{\lambda_0'\lambda_1\lambda_2'}\,
\varrho_{\lambda_2\lambda_2'}\,,
\end{equation}
where $\rho_{\lambda_0\lambda_0'}$ is the spin-density matrix of the emitting fermion line and $\varrho_{\lambda_2\lambda_2'}$ is the spin-density matrix of the emitted vector boson. If polarisation of the emitted boson is not tracked, one sets $\varrho_{\lambda_2\lambda_2'}=\delta_{\lambda_2\lambda_2'}$. In the diagonal approximation, $\varrho_{\lambda_2\lambda_2'}=\varrho_{\lambda_2\lambda_2}\,\delta_{\lambda_2\lambda_2'}$. For the purposes of analytic comparison, it is convenient to separate the transverse and longitudinal contributions,
\begin{equation}
F_T \;=\; \sum_{\lambda_2=\pm}\,F_{\lambda_2\lambda_2}\,,
\qquad
F_L^{(\mathrm{X})}\;=\;F_{0_*0_*}^{(\mathrm{X})}\,,
\qquad
F^{(\mathrm{X})}=F_T+F_L^{(\mathrm{X})}\,,
\end{equation}
where $0_*$ denotes the longitudinal state used in the shower (\emph{i.e.}\ the subtracted longitudinal current, or its GI completion), and where $F_{\lambda_2\lambda_2}$ stands for the corresponding helicity-diagonal contribution after summing $\lambda_0,\lambda_0',\lambda_1$ with $\rho$.
The relevant Ward identity implies, schematically,
\begin{equation}
p_\mu\,\mathcal{M}^\mu(\ldots V(p)\ldots)=M_V\,\mathcal{M}(\ldots G_V(p)\ldots)\,,
\label{eq:WI-schematic}
\end{equation}
up to our phase and coupling conventions.
Accordingly, the GI longitudinal splitting current is obtained by supplementing the subtraction remainder by the corresponding Goldstone-matching term implied by \eqref{eq:WI-schematic}.

Table~\ref{tab:qqV-helicity} summarises the non-vanishing reduced amplitudes in the notation of Ref.~\cite{Masouminia:2021kne}, with the SL prescription corresponding to the pure subtracted longitudinal current and the GI prescription obtained by adding the uniquely fixed Goldstone-matching contribution implied by the broken-theory Ward identity (cf.\ Sec.~\ref{sec:formalism}).  Concretely, the GI completion modifies \emph{only} the longitudinal entries ($\lambda_2=0_*$), while leaving the transverse helicity amplitudes unchanged; the additional terms are proportional to the Goldstone-fermion couplings
\begin{equation}
\mathcal{O}[G_V]\;\propto\;\bar q'(q_1)\,\big(y_L P_L + y_R P_R\big)\,q(q_0)\,G_V(q_2)\,,
\label{eq:WI-YC}
\end{equation}
and a channel-dependent coefficient $c_G$ (encoding the normalisation/convention for the Goldstone insertion in the matching). At the amplitude level, the GI completion can be formulated as:
\begin{equation}
\widehat H^{(\mathrm{GI})}_{\lambda_0\lambda_1 0_*} = 
\widehat H^{(\mathrm{SL})}_{\lambda_0\lambda_1 0_*} 
+ c_G \widehat H^{(\mathrm{G_V})}_{\lambda_0\lambda_1 0_*},
\end{equation}
with $\widehat H^{(\mathrm{G_V})}_{\lambda_0\lambda_1}$ built from the scalar (Goldstone) Yukawa current, i.e. an operator of the type shown in Eq.~\eqref{eq:WI-YC}.

\begin{table}[t]
\centering
\renewcommand{\arraystretch}{1.25}
\resizebox{0.8\textwidth}{!}{%
\begin{tabular}{|c c c || c | c|}
\hline
$\lambda_0$ & $\lambda_1$ & $\lambda_2$ &
\multicolumn{2}{c|}{$\widehat H^{\mathrm{SL}}_{\lambda_0\lambda_1\lambda_2} = \widehat H^{\mathrm{GI}}_{\lambda_0\lambda_1\lambda_2}$} 
\\
\hline \hline
$-$ & $-$ & $-$ &
\multicolumn{2}{c|}{$-\dfrac{\sqrt{2}\,g_L\,p_\perp}{(1-z)\sqrt{z}}\dfrac{e^{i\varphi}}{\mathcal{D}}$}
\\
$-$ & $-$ & $+$ &
\multicolumn{2}{c|}{$\dfrac{\sqrt{2z}\,g_L\,p_\perp}{1-z}\dfrac{e^{-i\varphi}}{\mathcal{D}}$}
\\[1mm]
$+$ & $+$ & $-$ &
\multicolumn{2}{c|}{$-\dfrac{\sqrt{2z}\,g_R\,p_\perp}{1-z}\dfrac{e^{i\varphi}}{\mathcal{D}}$}
\\
$+$ & $+$ & $+$ &
\multicolumn{2}{c|}{$\dfrac{\sqrt{2}\,g_R\,p_\perp}{(1-z)\sqrt{z}}\dfrac{e^{-i\varphi}}{\mathcal{D}}$}
\\[1mm]
$-$ & $+$ & $+$ &
\multicolumn{2}{c|}{$-\dfrac{\sqrt{2}}{\sqrt{z}}\big(g_L z\,m_0 - g_R m_1\big)\dfrac{1}{\mathcal{D}}$}
\\
$+$ & $-$ & $-$ &
\multicolumn{2}{c|}{$\dfrac{\sqrt{2}}{\sqrt{z}}\big(g_R z\,m_0 - g_L m_1\big)\dfrac{1}{\mathcal{D}}$}
\\
\hline
\hline
$\lambda_0$ & $\lambda_1$ & $\lambda_2$ &
$\widehat H^{\mathrm{SL}}_{\lambda_0\lambda_1\lambda_2}$ &
$\widehat H^{\mathrm{GI}}_{\lambda_0\lambda_1\lambda_2}$
\\
\hline \hline
$-$ & $-$ & $0_*$ &
$-\dfrac{2\,g_L\,\sqrt{z}\,m_2}{1-z}\dfrac{1}{\mathcal{D}}$ &
$\Big[-\dfrac{2\,g_L\,\sqrt{z}\,m_2}{1-z}
\;+\;
\dfrac{\sqrt{2}\,c_G\,\big(z\,m_0 y_R + m_1 y_L\big)}{\sqrt{z}}\Big]\dfrac{1}{\mathcal{D}}$
\\
$+$ & $+$ & $0_*$ &
$-\dfrac{2\,g_R\,\sqrt{z}\,m_2}{1-z}\dfrac{1}{\mathcal{D}}$ &
$\Big[-\dfrac{2\,g_R\,\sqrt{z}\,m_2}{1-z}
\;+\;
\dfrac{\sqrt{2}\,c_G\,\big(z\,m_0 y_L + m_1 y_R\big)}{\sqrt{z}}\Big]\dfrac{1}{\mathcal{D}}$
\\
$-$ & $+$ & $0_*$ &
$0$ &
$-\dfrac{\sqrt{2}\,c_G\,y_L\,p_\perp}{\sqrt{z}}\dfrac{e^{-i\varphi}}{\mathcal{D}}$
\\
$+$ & $-$ & $0_*$ &
$0$ &
$-\dfrac{\sqrt{2}\,c_G\,y_R\,p_\perp}{\sqrt{z}}\dfrac{e^{i\varphi}}{\mathcal{D}}$
\\
\hline
\end{tabular}}
\caption{Non-vanishing reduced helicity amplitudes for $q\to q'V$ in the SL and GI prescriptions.  Entries common to both prescriptions are grouped in the upper block, while the lower block collects the longitudinal ($0_*$) components, for which the GI completion adds Goldstone-matching terms proportional to $c_G$ and $(y_L,y_R)$.}
\label{tab:qqV-helicity}
\end{table}

Using the decomposition $F^{(\mathrm{X})}=F_T+F_L^{(\mathrm{X})}$ and performing the helicity sums implied by the shower construction, weighted by the diagonal entries of the spin-density matrix $\rho$, one obtains the following closed forms for the transverse contribution and for the longitudinal contribution in the SL and GI prescriptions. Writing the kernels as functions of $(z,\tilde q^2)$ (with $\tilde q$ the evolution variable of Sec.~\ref{sec:kinematics} and with $p_\perp^2$ eliminated using Eq.~\eqref{eq:hw7-pt2}), the transverse part is common to both schemes,
\begin{align}
F_T^{q\to q'V} =
{1 \over 2 \tilde{q}^2 z (1-z)} &\Bigg[  
    {\big( g_R^2 \; \rho_{++} + g_L^2 \;\rho_{--} \big) \over 2}
    \big(
    m_0^2 (1-z^2) - m_1^2 (1-z^2) - m_2^2 (1+z^2)
    \big)
    \nonumber \\
    &+ \big( g_R^2 \; \rho_{++} + g_L^2 \; \rho_{--} \big)\,\tilde{q}^2 z (1+z^2)
    + 4\,g_R g_L\,m_0 m_1\,\big(\rho_{--}+\rho_{++}\big)
\Bigg],
\label{eq:qqV-FT}
\end{align}
whereas the SL contribution reads
{\allowdisplaybreaks
\begin{align}
F_{L;\rm{SL}}^{q\to q'V} = 
{m_2^2 \over \tilde{q}^2 (1-z)^3}\,\big( g_R^2 \; \rho_{++}  + g_L^2 \;\rho_{--}  \big),
\label{eq:qqV-FL-SL}
\end{align}
while the GI longitudinal contribution becomes
\begin{align}
F_{L;\rm{GI}}^{q\to q'V} &=
{1 \over \tilde{q}^2 (1-z)^3}\,\big( g_R^2 \;\rho_{++} + g_L^2 \; \rho_{--} \big)\, m_2^2
\nonumber\\
&\quad - {\sqrt{2}\;c_G\; m_2 \over \tilde{q}^2 z (1-z)^2}
\Big[
\big(g_R \;\rho_{++}\, y_L + g_L \;\rho_{--}\, y_R\big)\, m_0 z
+
\big(g_R \;\rho_{++}\, y_R + g_L \;\rho_{--}\, y_L\big)\, m_1
\Big]
\nonumber\\
&\quad + {c_G^2 \over 2 \tilde{q}^2 z (1-z)}
\Bigg[
\big( \rho_{--}\,y_L^2 + \rho_{++}\,y_R^2 \big)
\big( m_0^2 + m_1^2 - m_2^2 + \tilde{q}^2 z (1-z)^2 \big)
\nonumber\\
&\hspace{1.1in} + 4\,m_0 m_1\,y_R y_L\,\big(\rho_{--}+\rho_{++}\big)
\Bigg]
.
\label{eq:qqV-FL-GI}
\end{align}}

Equations \eqref{eq:qqV-FT}-\eqref{eq:qqV-FL-GI} make explicit the structural point emphasised in Sec.~\ref{sec:formalism}: the SL and GI prescriptions share an identical transverse sector, while their longitudinal sectors differ by terms controlled by the Goldstone-matching, including both interference ($\propto c_G$) and pure Goldstone ($\propto c_G^2$) contributions, which are parametrically suppressed for light fermions in Standard Model applications but become essential whenever symmetry-breaking effects and Yukawa insertions compete with the naive $p^\mu/M_V$ subtraction.

For the purpose of understanding the scaling behaviour and for cross-checks against known limits, it is useful to decompose the GI kernel into a massless piece (defined by formally setting $m_0=m_1=m_2=0$ at fixed $z$) and a remainder.  One finds
\begin{equation}
F^{q\to q'V}_{\mathrm{GI,massless}}(z)
=
\Big(g_L^2 \; \rho_{--} +g_R^2\;\rho_{++}\Big)\frac{1+z^2}{1-z}
+
c_G^2\Big(y_L^2\;\rho_{--}+y_R^2\;\rho_{++}\Big)\frac{(1-z)^2}{1-z}\,,
\label{eq:qqV-GI-massless}
\end{equation}
with the corresponding massive remainder defined as
\begin{equation}
F^{q\to q'V}_{\mathrm{GI, massive}}(z,\tilde q^2)
\;\equiv\;
F^{q\to q'V}_{\mathrm{GI}}(z,\tilde q^2)-F^{q\to q'V}_{\mathrm{GI,massless}}(z).
\label{eq:qqV-GI-massive-def}
\end{equation}
In particular, Eq.~\eqref{eq:qqV-GI-massless} makes transparent that the GI completion introduces a genuinely new $(1-z)$-suppressed contribution already in the kinematic massless limit (neglecting mass terms in the kernels), originating from the Goldstone sector and absent in the pure SL, while Eq.~\eqref{eq:qqV-FL-GI} shows how this term is continuously connected to the massive, ultracollinear regime through the Ward-identity matching.

\begin{figure} [h]
\centering
% J/ψ
\includegraphics[width=.49\textwidth]{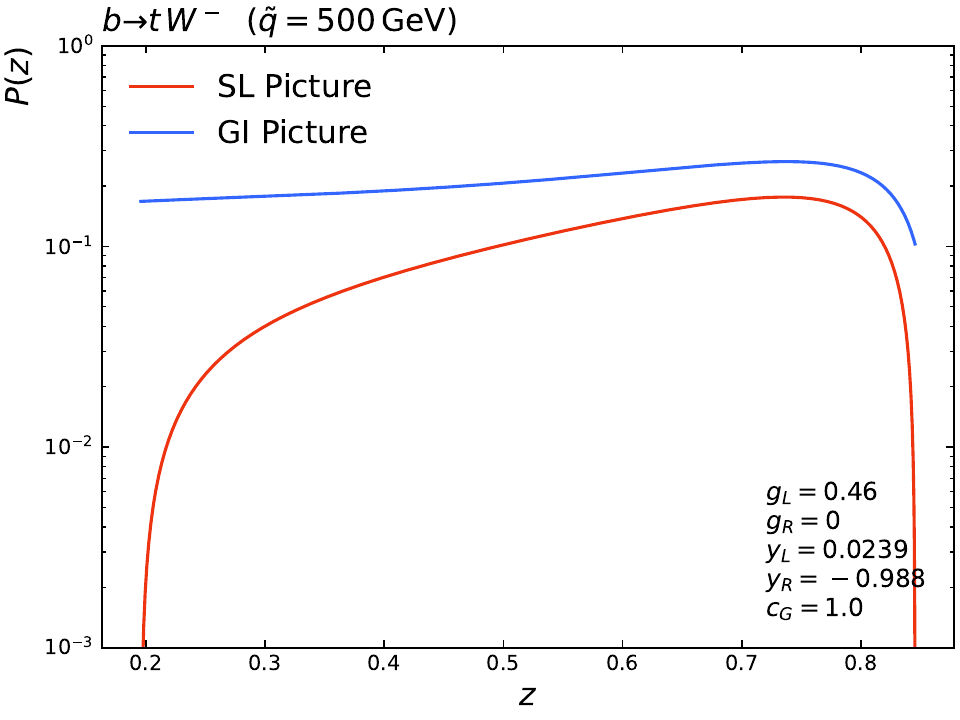}
\includegraphics[width=.49\textwidth]{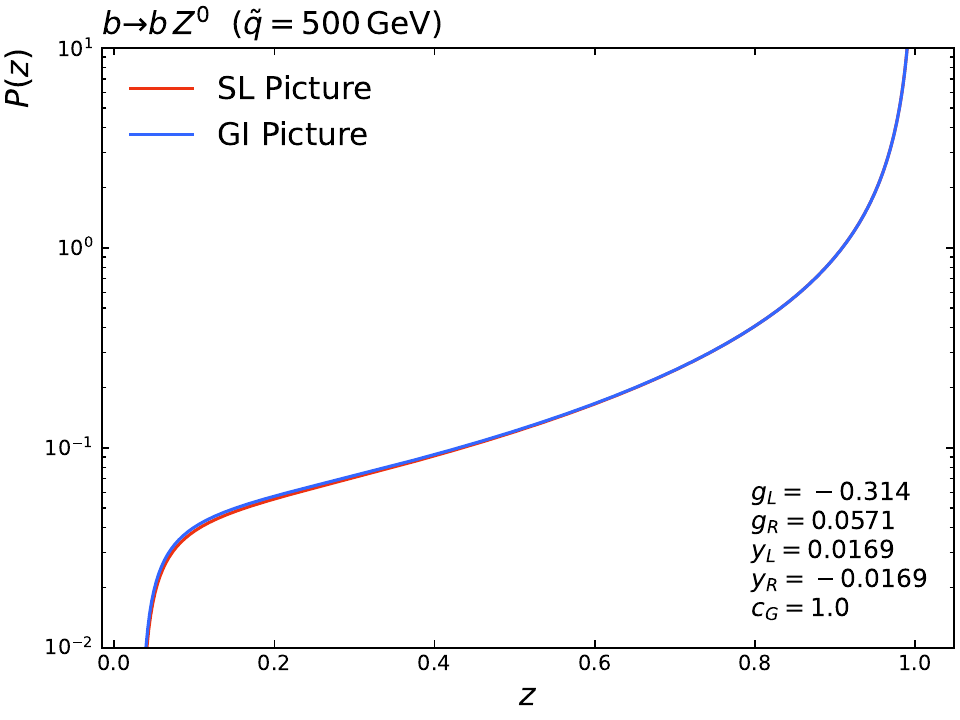}
\caption{\small
Spin-averaged quasi-collinear splitting kernels $P(z)$ for representative $q\to q'V$ branchings at fixed transverse scale $\tilde{q}=500~\mathrm{GeV}$, comparing the original SL prescription (red) with the GI completion (blue). Left: $b\to t\,W^-$, for which the Ward-identity (Goldstone) matching terms are Yukawa-enhanced through the heavy-quark mass and induce a visible deformation of the longitudinal contribution over a broad range of $z$. Right: $b\to b\,Z^0$, where the Goldstone couplings are parametrically suppressed, and the two prescriptions are practically indistinguishable at the level of the inclusive kernel.}
\label{fig:qqv_compare}
\end{figure}

Figure~\ref{fig:qqv_compare} illustrates, at the level of the spin-averaged $q\to q'V$ kernel, how the GI completion reorganises the longitudinal sector relative to the SL prescription while leaving the transverse contribution unchanged.  For $b\to t\,W^-$ (left), the GI longitudinal amplitude contains, in addition to the finite subtraction remainder, the Ward-identity-mandated Goldstone-matching term, whose interference with the vector remainder and whose pure-Goldstone contribution are both controlled by Yukawa couplings; the resulting longitudinal piece is therefore sensitive to the heavy-quark masses and can shift the inclusive kernel by an $\mathcal{O}(1)$ amount across intermediate $z$, even at $\tilde{q}=500~\mathrm{GeV}$, where pure mass-power effects are otherwise expected to decouple.  By contrast, for $b\to b\,Z^0$ (right), the associated Goldstone couplings are suppressed by $m_b/M_W$ and the GI correction is correspondingly negligible, so that the SL and GI kernels coincide to very high accuracy once summed over polarisations.  In both channels, the strong rise towards $z\to1$ reflects the endpoint enhancement intrinsic to the quasi-collinear kinematics used by the shower, and the plots therefore provide a clean, channel-by-channel diagnostic of where the longitudinal completion can be phenomenologically relevant before embedding the kernels into Sudakov evolution.

\subsection{$V\to V'V''$ splittings}
\label{sec:VVV}

We consider the generic electroweak branching
\begin{equation}
V_0(q_0,\lambda_0)\;\longrightarrow\; V_1(q_1,\lambda_1)\;+\;V_2(q_2,\lambda_2)\,,
\end{equation}
with gauge vector bosons and helicities $\lambda_i=\pm,0_*$. The relevant Lorentz structure is the standard three-vector
vertex,
\begin{equation}
\Gamma^{\mu\nu\rho}(q_0,q_1,q_2)
=
g^{\mu\nu}(q_0-q_1)^{\rho}
+
g^{\nu\rho}(q_1-q_2)^{\mu}
+
g^{\rho\mu}(q_2-q_0)^{\nu},
\label{eq:vvv-vertex}
\end{equation}
so that the reduced splitting amplitudes are built from
\begin{equation}
\widehat H_{\lambda_0\lambda_1\lambda_2}
=
g_{012}\;
\varepsilon_\mu(q_0,\lambda_0)\,
\varepsilon_\nu^\ast(q_1,\lambda_1)\,
\varepsilon_\rho^\ast(q_2,\lambda_2)\,
\Gamma^{\mu\nu\rho}(q_0,q_1,q_2),
\label{eq:vvv-Hdef}
\end{equation}
with $g_{012}$ the channel-dependent triple-gauge coupling
(e.g.\ $g_{ZWW}=g\,c_W$ in the SM). As for $q\to q'V$, the \textsf{Herwig}
spin-dependent kernel is obtained by contracting $\widehat H$ with the parent
spin-density matrix $\rho$ and, if desired, retaining a spin-density matrix
for the emitted vectors for subsequent evolution/decay \cite{Masouminia:2021kne}. 
In the GI prescription, the longitudinal shower state $0_*$ is understood as the
subtracted longitudinal current supplemented by the Ward-identity
(Goldstone) completion. Concretely, for $V_0\to V_1V_2$ we implement this at the
level of reduced amplitudes as
\begin{align}
\widehat H^{\mathrm{GI}}_{\lambda_0\lambda_1\lambda_2}
=
\widehat H^{\mathrm{SL}}_{\lambda_0\lambda_1\lambda_2}
+ c_G\Big[
\delta_{\lambda_0 0_*}\,\widehat H^{(G_0)}_{\lambda_1\lambda_2}
+ \delta_{\lambda_1 0_*}\,\widehat H^{(G_1)}_{\lambda_0\lambda_2}
+ \delta_{\lambda_2 0_*}\,\widehat H^{(G_2)}_{\lambda_0\lambda_1}
\Big],
\label{eq:vvv-GI-completion}
\end{align}
with the Goldstone-matching pieces given by the momentum contractions of the
three-vector vertex,
\begin{align}
\widehat H^{(G_0)}_{\lambda_1\lambda_2}
&\equiv
g_{012}\,\frac{\sqrt{2}}{m_0}\,
q_{0\mu}\,
\varepsilon_\nu^\ast(q_1,\lambda_1)\,
\varepsilon_\rho^\ast(q_2,\lambda_2)\,
\Gamma^{\mu\nu\rho}(q_0,q_1,q_2),\\
\widehat H^{(G_1)}_{\lambda_0\lambda_2}
&\equiv
g_{012}\,\frac{\sqrt{2}}{m_1}\,
\varepsilon_\mu(q_0,\lambda_0)\,
q_{1\nu}\,
\varepsilon_\rho^\ast(q_2,\lambda_2)\,
\Gamma^{\mu\nu\rho}(q_0,q_1,q_2),\\
\widehat H^{(G_2)}_{\lambda_0\lambda_1}
&\equiv
g_{012}\,\frac{\sqrt{2}}{m_2}\,
\varepsilon_\mu(q_0,\lambda_0)\,
\varepsilon_\nu^\ast(q_1,\lambda_1)\,
q_{2\rho}\,
\Gamma^{\mu\nu\rho}(q_0,q_1,q_2).
\end{align}

Explicitly, the spin-correlated kernel can be written in the same density-matrix form as for $q\to q'V$ splitting case,
\begin{equation}
F^{(\mathrm{X})}_{V\to V'V''}(z,\tilde q^2)\;\propto\;
\sum_{\lambda_0,\lambda_0',\lambda_1,\lambda_1',\lambda_2,\lambda_2'}
\rho_{\lambda_0\lambda_0'}\,
\widehat H^{(\mathrm{X})}_{\lambda_0\lambda_1\lambda_2}\,
\widehat H^{(\mathrm{X})\ast}_{\lambda_0'\lambda_1'\lambda_2'}\,
\varrho^{(1)}_{\lambda_1\lambda_1'}\,
\varrho^{(2)}_{\lambda_2\lambda_2'}\,,
\label{eq:vvv-rho-contract}
\end{equation}
where $\rho$ is the parent ($V_0$) spin-density matrix and $\varrho^{(1,2)}$ are the (optionally retained) spin-density matrices for $V_{1,2}$; for inclusive splitting probabilities one takes $\varrho^{(1)}=\varrho^{(2)}=\delta$ and traces over the daughter polarisations.
Because all three legs carry spin~1, it is useful to decompose the emission
probability into contributions according to whether each of
$(\lambda_0,\lambda_1,\lambda_2)$ is transverse ($T\equiv\pm$) or longitudinal
($L\equiv 0_*$). Denoting the longitudinal state used in the shower by $0_*$
(as in Sec.~\ref{sec:formalism}), we define
\begin{align}
F_{TTT} &= 
\sum_{\lambda_0\in T}\sum_{\lambda_0'\in T}\sum_{\lambda_1\in T}\sum_{\lambda_2\in T}
\rho_{\lambda_0\lambda_0'}\,
\widehat H_{\lambda_0\lambda_1\lambda_2}\,
\widehat H^\ast_{\lambda_0'\lambda_1\lambda_2},
\\
F_{TTL} &= 
\sum_{\lambda_0\in T}\sum_{\lambda_0'\in T}\sum_{\lambda_1\in T}
\rho_{\lambda_0\lambda_0'}\,
\widehat H_{\lambda_0\lambda_1 0_*}\,
\widehat H^\ast_{\lambda_0'\lambda_1 0_*},
\\
F_{TLT} &= 
\sum_{\lambda_0\in T}\sum_{\lambda_0'\in T}\sum_{\lambda_2\in T}
\rho_{\lambda_0\lambda_0'}\,
\widehat H_{\lambda_0 0_* \lambda_2}\,
\widehat H^\ast_{\lambda_0' 0_* \lambda_2},
\\
F_{LTT} &= 
\sum_{\lambda_1\in T}\sum_{\lambda_2\in T}
\rho_{0_*0_*}\,
\widehat H_{0_*\lambda_1\lambda_2}\,
\widehat H^\ast_{0_*\lambda_1\lambda_2},
\end{align}
and analogously $F_{TLL}$, $F_{LTL}$, $F_{LLT}$ and $F_{LLL}$. In an unpolarised shower $\rho$ is diagonal and $\varphi$-averaging removes interference terms between different helicity states; nevertheless, we keep the polarisation-resolved definitions explicit, since they are required by the full spin-correlation machinery. The remaining helicity sectors vanish at $\mathcal{O}(\lambda^0)$ in the quasi-collinear expansion with the shower longitudinal basis $0^\ast$.

Table~\ref{tab:vvv-helicity} summarises the non-vanishing reduced amplitudes in the aligned frame in the two longitudinal prescriptions. The SL picture uses only the subtracted longitudinal current ($0_*$)
without explicit Goldstone-matching. The GI completion
implements the Ward-identity completion of the longitudinal sector in the broken
theory. Concretely, the transverse block ($\lambda=\pm$) is unchanged, while
entries involving $0_*$ differ by fixed mass- and $z$-dependent terms.

For compactness, we define the dimensionless ratios
\begin{equation}
\mu_i \equiv \frac{m_i}{\sqrt{t}},
\qquad
\kappa \equiv \sqrt{1+\mu_0^2-\frac{\mu_1^2}{z}-\frac{\mu_2^2}{1-z}}
\;=\;\frac{p_\perp}{\sqrt{t \; z(1-z)}}\,,
\label{eq:vvv-kappa}
\end{equation}
and write the azimuthal phases as $e^{\pm i\varphi}$.

\begin{table}[t]
\centering
\renewcommand{\arraystretch}{0.95}
\resizebox{0.7\textwidth}{!}{%
\begin{tabular}{|c c c || c | c|}
\hline
$\lambda_0$ & $\lambda_1$ & $\lambda_2$ &
\multicolumn{2}{c|}{$\widehat H^{\mathrm{SL}}_{\lambda_0\lambda_1\lambda_2} = \widehat H^{\mathrm{GI}}_{\lambda_0\lambda_1\lambda_2}$}
\\
\hline \hline
$-$ & $-$ & $-$ &
\multicolumn{2}{c|}{$g_{012}\;e^{i\varphi}\;\dfrac{\kappa}{\sqrt{z(1-z)}}$}
\\
$-$ & $-$ & $+$ &
\multicolumn{2}{c|}{$-g_{012}\;e^{-i\varphi}\;\kappa\sqrt{\dfrac{z}{1-z}}$}
\\
$-$ & $+$ & $-$ &
\multicolumn{2}{c|}{$-g_{012}\;(1-z)\;e^{-i\varphi}\;\kappa\sqrt{\dfrac{1-z}{z}}$}
\\
$+$ & $-$ & $+$ &
\multicolumn{2}{c|}{$g_{012}\;(1-z)\;e^{i\varphi}\;\kappa\sqrt{\dfrac{1-z}{z}}$}
\\
$+$ & $+$ & $-$ &
\multicolumn{2}{c|}{$g_{012}\;e^{i\varphi}\;\kappa\sqrt{\dfrac{z}{1-z}}$}
\\
$+$ & $+$ & $+$ &
\multicolumn{2}{c|}{$-g_{012}\;e^{-i\varphi}\;\dfrac{\kappa}{\sqrt{z(1-z)}}$}
\\[1mm]
$-$ & $-$ & $0_*$ &
\multicolumn{2}{c|}{$\sqrt{2}\,g_{012}\;\mu_2\;\dfrac{z}{1-z}$}
\\
$-$ & $0_*$ & $-$ &
\multicolumn{2}{c|}{$-\sqrt{2}\,g_{012}\;\mu_1\;\dfrac{1-z}{z}$}
\\
$+$ & $0_*$ & $+$ &
\multicolumn{2}{c|}{$-\sqrt{2}\,g_{012}\;\mu_1\;\dfrac{1-z}{z}$}
\\
$+$ & $+$ & $0_*$ &
\multicolumn{2}{c|}{$\sqrt{2}\,g_{012}\;\mu_2\;\dfrac{z}{1-z}$}
\\
\hline
\hline
$\lambda_0$ & $\lambda_1$ & $\lambda_2$ &
$\widehat H^{\mathrm{SL}}_{\lambda_0\lambda_1\lambda_2}$ &
$\widehat H^{\mathrm{GI}}_{\lambda_0\lambda_1\lambda_2}$
\\
\hline \hline
$0_*$ & $-$ & $+$ &
$-\sqrt{2}\,g_{012}\;\mu_0\;(1-z)$ &
$+\sqrt{2}\,g_{012}\;\big(\mu_1-\mu_0(1-z)\big)$
\\
$0_*$ & $+$ & $-$ &
$-\sqrt{2}\,g_{012}\;\mu_0\;(1-z)$ &
$+\sqrt{2}\,g_{012}\;\big(\mu_1-\mu_0(1-z)\big)$
\\
$-$ & $+$ & $0_*$ &
$0$ &
$+\sqrt{2}\,g_{012}\;\mu_2$
\\
$-$ & $0_*$ & $+$ &
$0$ &
$+\sqrt{2}\,g_{012}\;\mu_1$
\\
$+$ & $0_*$ & $-$ &
$0$ &
$+\sqrt{2}\,g_{012}\;\mu_1$
\\
$+$ & $-$ & $0_*$ &
$0$ &
$+\sqrt{2}\,g_{012}\;\mu_2$
\\
\hline
\end{tabular}}
\caption{Non-vanishing reduced helicity amplitudes for $V_0\to V_1V_2$ in the SL and GI prescriptions, using the \textsf{Herwig~7} phase convention $\varphi$ and the dimensionless ratios $\mu_i=m_i/\sqrt{t}$ and $\kappa$ defined in Eq.~\eqref{eq:vvv-kappa}. Entries common to both prescriptions are grouped in the upper block; differences are confined to the lower block and involve the longitudinal state $0_*$.}
\label{tab:vvv-helicity}
\end{table}

Performing the helicity sums implied by the shower construction (with $\rho$
kept diagonal), and with the kinematic substitutions used in our analytic
derivation, the transverse sector is identical in both prescriptions:
\begin{align}
F_{TTT}^{\mathrm{SL}}(z,\tilde q^2)
&=
F_{TTT}^{\mathrm{GI}}(z,\tilde q^2)
 \\
&= (\rho_{--}+\rho_{++})
\frac{2\big(1-z+z^2\big)^2}{\tilde q^2\,z^3\,(z-1)^3}
\Big[
\tilde q^2\,z^2(z-1)^2 + m_0^2 \, z (z-1) -m_1^2\,(1-z)-m_2^2 \, z
\Big]
. \nonumber
\label{eq:vvv-FTTT}
\end{align}
By contrast, the sectors with a single longitudinal leg differ between SL
and GI. For the configuration with $V_2$ longitudinal, one finds
\begin{align}
F_{TTL}^{\mathrm{SL}}(z,\tilde q^2)
&=2\,m_2^2\,\left(\rho_{--}+\rho_{++}\right)
\frac{z}{\tilde q^2\,(1-z)^3}\,,
\\
F_{TTL}^{\mathrm{GI}}(z,\tilde q^2)
&=2\,m_2^2\,\left(\rho_{--}+\rho_{++}\right) \bigg[
\frac{z}{\tilde q^2\,(1-z)^3}
+
\frac{c_G^2}{\tilde q^2\,(1-z) z}\,
\bigg]
,
\end{align}
while for the configuration with $V_1$ longitudinal,
\begin{align}
F_{TLT}^{\mathrm{SL}}(z,\tilde q^2)
&=2\,m_1^2\,\left(\rho_{--}+\rho_{++}\right)
\frac{1-z}{\tilde q^2\,z^3}\,,
\label{eq:vvv-FTLT-daw}
\\
F_{TLT}^{\mathrm{GI}}(z,\tilde q^2)
&=2\,m_1^2\,\left(\rho_{--}+\rho_{++}\right) \bigg[
\frac{1-z}{\tilde q^2\,z^3}
+ \frac{c_G^2}{\tilde q^2\,(1-z)\,z}
\bigg],
\label{eq:vvv-FTLT-gi}
\end{align}
and for a longitudinal parent ($V_0$ longitudinal),
\begin{align}
F_{LTT}^{\mathrm{SL}}(z,\tilde q^2)
&=4\,m_0^2\,\rho_{00}\,
\frac{1-z}{\tilde q^2\,z}\,,
\label{eq:vvv-FLTT-daw}
\\
F_{LTT}^{\mathrm{GI}}(z,\tilde q^2)
&=4\,m_0^2\,\rho_{00}\,\bigg[
\frac{1-z}{\tilde q^2\,z}
+ \frac{c_G\;(c_G + 2 z - 2)}{\tilde q^2\;(1-z)\,z}
\bigg].
\label{eq:vvv-FLTT-gi}
\end{align}
All remaining polarisation sectors ($TLL$, $LTL$, $LLT$, $LLL$) vanish in
this quasi-collinear construction with the shower longitudinal basis, in
agreement with the explicit checks in our notebook derivation.

It is often useful to write the difference between the GI and SL
prescriptions directly. Summing the polarisation-resolved contributions into
the spin-summed splitting function $F_{(\mathrm{X})}(z,\tilde q^2)$, the net
shift induced by the GI completion can be expressed compactly as
\begin{align}
F_{\mathrm{GI}}^{V\to V'V''}(z,\tilde q^2)-F_{\mathrm{SL}}^{V\to V'V''}(z,\tilde q^2)
&=
  2\; c_G^2\; (\rho_{--}+\rho_{++}) \; \frac{\,(m_1^2+m_2^2)}{\tilde q^2\,(1-z)\,z}\,
\nonumber \\
&+ 2 \; c_G \; \rho_{00} \; \frac{m_0^2}{\tilde q^2\,(1-z)\,z} \big[ 2c_G - 4 (1-z) \big]
\; ,
\label{eq:vvv-delta}
\end{align}
showing explicitly that the two prescriptions coincide in the strict
massless/collinear limit and differ only by symmetry-breaking mass effects. Moreover, separating the GI result into a strict massless contribution and a massive
remainder (defined by formally setting $m_i\to0$ at fixed $z$), one finds
\begin{align}
F_{\mathrm{GI,\,massless}}^{V\to V'V''}(z)
&=
2\,(\rho_{--}+\rho_{++})\,
\frac{\big(1-z+z^2\big)^2}{z(1-z)}\,,
\label{eq:vvv-massless}
\\
F_{\mathrm{GI,\,massive}}^{V\to V'V''}(z,\tilde q^2)
&=
\frac{2\,(\rho_{--}+\rho_{++})}{\tilde q^2\,z^2(1-z)^2}
\Big[
(1-z+z^2)^2 \, m_0^2 - (1 - c_G^2 z - z^2 + c_G^2 z^2 + z^3) \, m_1^2 
\nonumber \\
&- (1 - z - c_G^2 z + 2 z^2 + c_G^2 z^2 - z^3) \, m_2^2
\Big]
+ 4 \; \rho_{00} \frac{m_0^2}{\tilde q^2\,(1-z)} \;(1 - c_G - z)^2.
\label{eq:vvv-massive}
\end{align}

\begin{figure}[t]
\centering
\includegraphics[width=.49\textwidth]{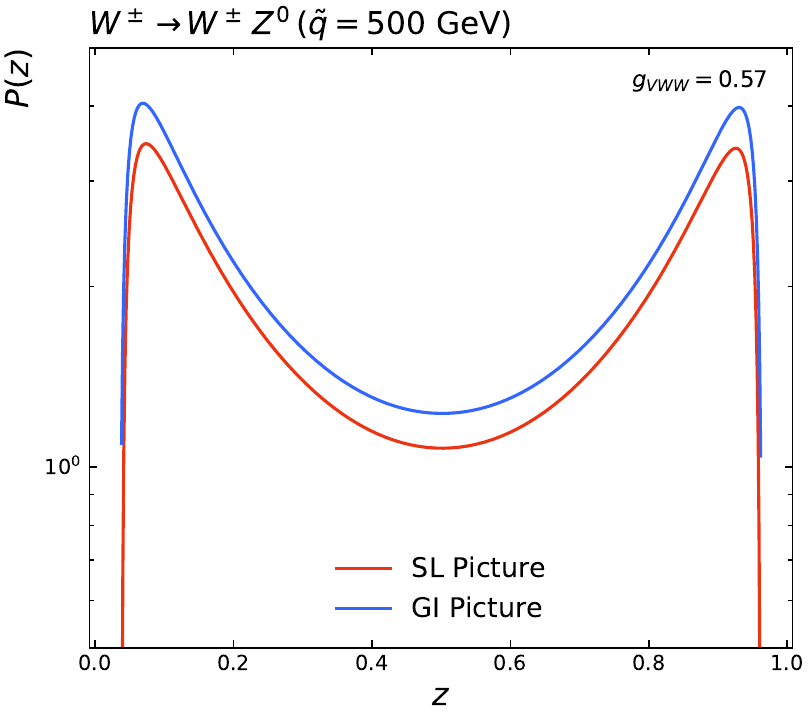}
\includegraphics[width=.49\textwidth]{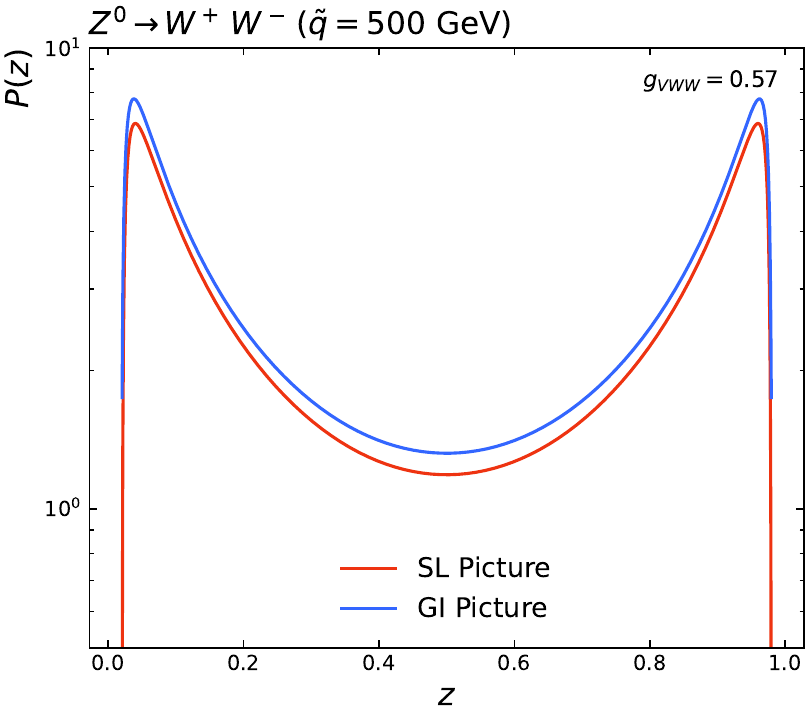}
\includegraphics[width=.49\textwidth]{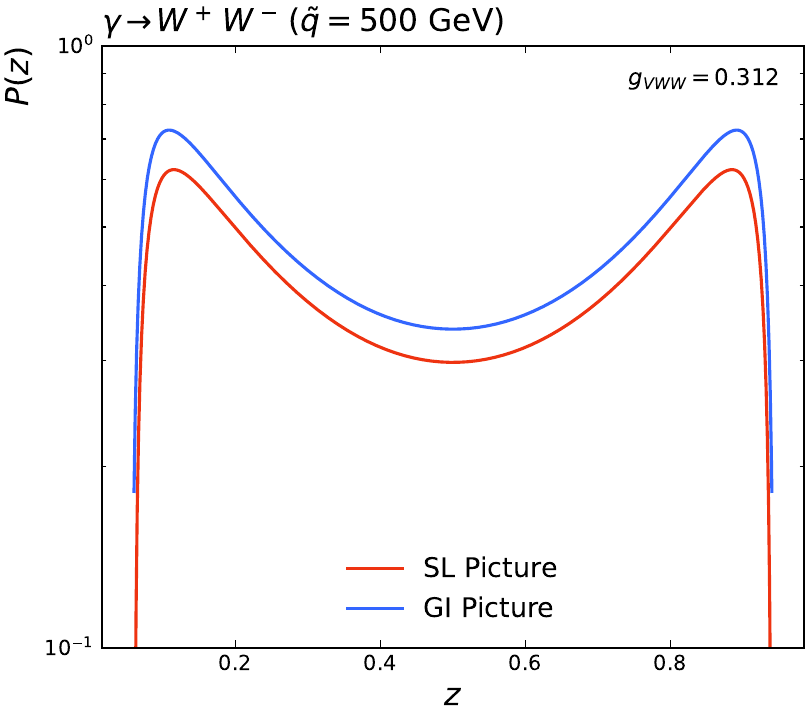}
\caption{\small Comparison of the analytic $V_0\to V_1V_2$ splitting kernels in
the SL and GI prescriptions for three representative
channels: $W\to WZ$ (left), $Z\to W^+W^-$ (middle) and $\gamma\to W^+W^-$ (right).
In each case we show the spin-summed kernel $F^{(\mathrm{X})}(z,\tilde q^2)$ as
a function of the lightcone momentum fraction $z$ at fixed evolution scale
$\tilde q$ (with pole masses for the external vectors).}
\label{fig:vvv_compare}
\end{figure}

Figure~\ref{fig:vvv_compare} illustrates that the impact of the GI
completion in bosonic splittings is highly structured: the SL and GI
pictures share an identical $TTT$ contribution, but differ whenever exactly one
daughter is in the longitudinal shower state $0_*$. The net effect is a
channel-independent, symmetry-breaking shift proportional to $(m_1^2+m_2^2)$,
Eq.~\eqref{eq:vvv-delta}, which is enhanced in asymmetric branchings through
the $1/[z(1-z)]$ factor and is suppressed as $\tilde q$ becomes large compared
to the electroweak scale. This makes $V\to V'V''$ branchings a particularly
clean arena to isolate longitudinal systematics: any deviation between the two
schemes can be traced back to the Ward-identity completion of the single-$L$
sectors, rather than to the transverse dynamics, which is common to both
constructions.

\section{Results and Discussion}
\label{sec:res}

To disentangle genuine kernel-level effects from shower-evolution artefacts, we study single- and full-resummed predictions separately. In the single-resummed setup, we restrict the evolution to at most one electroweak emission, which isolates the quasi-collinear splitting operators (including the longitudinal prescriptions under comparison) from higher-multiplicity recoil chains and from changes in the Sudakov exponent induced by repeated emissions. The full-resummed evolution then reinstates the complete Markovian shower, where modified longitudinal pieces feed back non-linearly through the no-emission probability, emission ordering and phase-space depletion, and where differences can migrate between exclusive and inclusive observables. Presenting both levels, therefore, provides a controlled validation of the implementation at fixed splitting kinematics and a physically relevant assessment of its impact once exponentiated to all orders within the shower.

For both the single-resummed and full-resummed studies in this section, we use \textsf{Herwig}'s internal leading-order matrix elements, and keep the shower configuration identical between longitudinal prescriptions except for the choice of the longitudinal scheme flag (SL vs. GI). Unless stated otherwise, we analyse events with the \textsf{Herwig~7}-\textsf{Rivet} interface, running custom \textsf{Rivet}~\cite{Bierlich:2019rhm,Bierlich:2024vqo} analyses and writing histograms in \textsf{YODA}~\cite{Buckley:2023xqh} format for subsequent post-processing with the standard \textsf{YODA} tools.  In the single-resummed tests, we isolate the shower effects by switching off multi-parton interactions, hadronisation, and decays, and by limiting the evolution to a single final-state emission; electroweak gauge bosons are kept stable.  In the full-resummed, LHC-like runs, we enable the complete shower evolution (with the process-dependent settings given in the input files), retaining the same PDF choice, electroweak scheme, and kinematic cuts as in the corresponding single-emission setups.  All generator input files and \textsf{Rivet} analysis routines used to produce the figures in this section are provided alongside this paper.

\subsection{Single-Resummed Evolution}
\label{sec:srs}

A further, and qualitatively distinct, validation step is to compare the single-emission resummed predictions obtained in the SL and GI prescriptions.  Fixed-order comparisons, as already presented for the SL implementation in Ref.~\cite{Masouminia:2021kne}, probe the correctness of the quasi-collinear kernels at $\mathcal{O}(\alpha_{\mathrm{EW}})$, but they do not test the way in which those kernels are embedded into a Markovian shower evolution, i.e.\ the interplay between the helicity-dependent branching operator, the shower kinematics map, and the Sudakov exponentiation which enforces unitarity.  By restricting to a configuration in which at most one final-state electroweak emission is generated, one obtains a controlled observable bridge between fixed-order and the fully iterated shower: the emission probability is still governed by the resummed (Sudakov-suppressed) single-branching functional, yet it remains sufficiently simple that differences can be traced directly to the longitudinal construction rather than to higher-multiplicity histories, competing emissions, hadronisation, or decay modelling.  This is particularly relevant here because the SL and GI prescriptions coincide identically in the transverse sector and differ only through the longitudinal completion implied by broken-theory Ward identities; a single-resummed test therefore isolates whether these longitudinal differences survive the shower embedding in a physically meaningful way, and whether they appear in the expected channels and kinematic regions (notably those sensitive to symmetry-breaking and Yukawa structures) before turning to fully exclusive multi-emission phenomenology.

\begin{figure}[t]
\centering
\includegraphics[width=.49\textwidth]{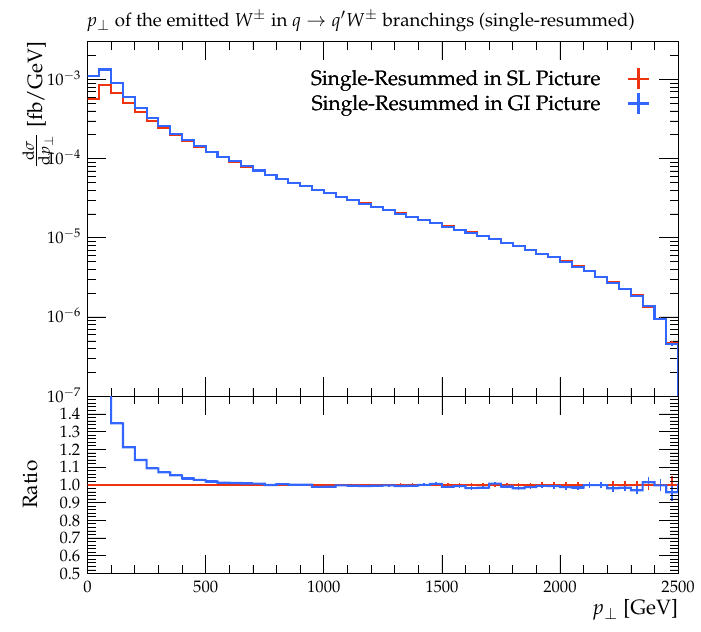}
\includegraphics[width=.49\textwidth]{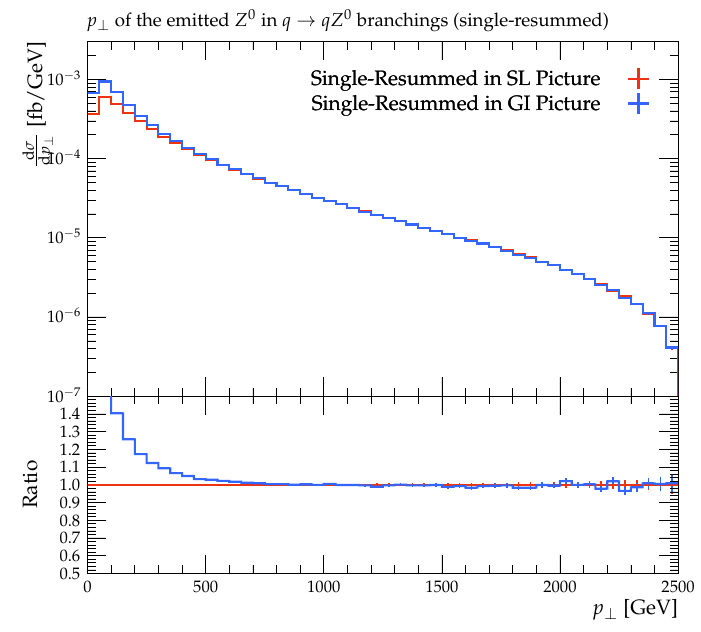}
\includegraphics[width=.49\textwidth]{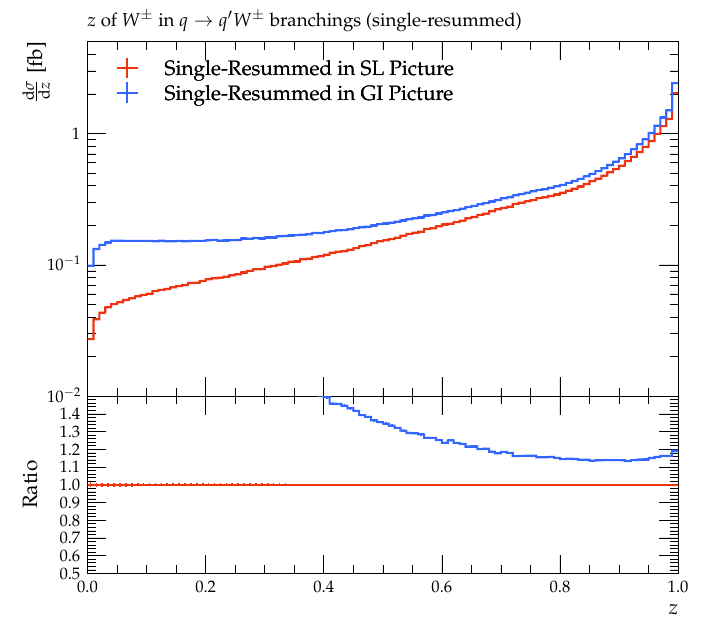}
\includegraphics[width=.49\textwidth]{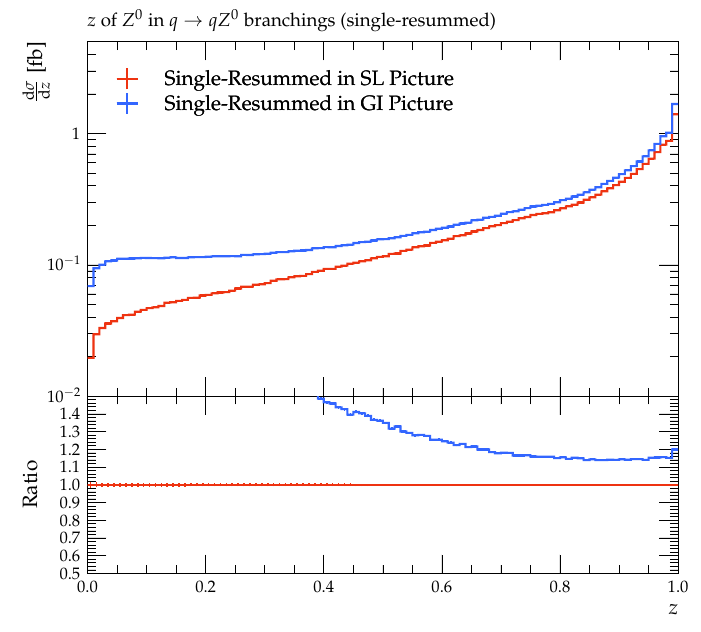}
\caption{\small
Single-resummed final-state radiation (SRS) comparison between the SL and GI longitudinal prescriptions for $q\to q'V$ branchings in a clean $e^+e^-\to q\bar q$ setup at $\sqrt{s}=10~\mathrm{TeV}$ (no hadronisation and no decays; exactly one electroweak gauge boson in the event record). Left column: charged-current emission ($q\to q'W^+$); right column: neutral-current emission ($q\to qZ^0$). Shown are the boson transverse-momentum spectrum $p_{\perp,V}$ (top) and the lightcone momentum fraction proxy $z$ used in the shower mapping (bottom); ratio panels display GI/SL.}
\label{fig:qqv_srs}
\end{figure}

Figure~\ref{fig:qqv_srs} demonstrates, in a deliberately controlled single-emission Sudakov setting, how the GI completion propagates beyond the level of analytic kernels into a shower-embedded prediction. In both $q\to q'W^+$ and $q\to qZ^0$, the SL and GI results remain close over most of the kinematic range, with a smooth, predominantly shape-level scheme dependence that is most visible in the soft/collinear region of the $p_{\perp,V}$ spectrum and in the associated redistribution in the mapped energy-sharing variable $z$. The charged-current channel exhibits the larger response, as expected for longitudinal-sensitive emissions, while the neutral-current case remains more weakly affected overall. The significance of this test is twofold: first, it provides a direct diagnostic that the GI longitudinal completion is implemented consistently within the \textsf{Herwig} kinematics map and unitarised evolution (since the two prescriptions coincide in the transverse sector), and second, it establishes that any phenomenological differences observed in fully exclusive simulations can be interpreted as originating from longitudinal systematics in the electroweak shower rather than from secondary modelling choices such as multi-emission competition, hadronisation, or unstable-particle decays.

It is worth stressing that an apparent inversion between the size of the analytic splitting kernels and the size of the observed shower-level signatures is not only possible but, in an exclusive single-emission study, entirely expected.  At the level of the branching operator, one compares local quasi-collinear kernels $P^{(\mathrm{X})}(t,z)$ (with $\mathrm{X}=\mathrm{SL},\mathrm{GI}$), whereas the event-level distributions in Fig.~\ref{fig:qqv_srs} are generated by a unitarised Markov evolution and are therefore weighted by Sudakov factors.  Schematically, the probability density for producing exactly one final-state EW emission at evolution scale $t$ can be written as
\begin{equation}
\mathrm{d}\mathcal{P}_1^{(\mathrm{X})}(t,z)
\;=\;
\Delta^{(\mathrm{X})}(Q,t)\;
\Big[\mathrm{d}t\,\mathrm{d}z\;P^{(\mathrm{X})}(t,z)\Big]\;
\Delta^{(\mathrm{X})}(t,t_{\mathrm{cut}})\,,
\label{eq:one-emission-sudakov}
\end{equation}
where $Q$ is the starting scale and $t_{\mathrm{cut}}$ the shower cutoff.  The Sudakov form factor
\begin{equation}
\Delta^{(\mathrm{X})}(Q,t)
\;=\;
\exp\!\left[
-\int_{t}^{Q}\mathrm{d}t'\!\int\mathrm{d}z'\;P^{(\mathrm{X})}(t',z')
\right]
\label{eq:sudakov}
\end{equation}
is itself controlled by the integral of the same kernel. Consequently, a prescription (such as the GI completion) which yields a locally larger longitudinal contribution in parts of $(t,z)$ phase space typically also yields a \emph{stronger} Sudakov suppression, and it can enhance the probability for additional emissions.  Since our single-resummed validation explicitly targets a one-boson topology (either by construction through emission limits or, equivalently, through an analysis-level requirement of exactly one EW vector boson in the final state), the accepted sample is an \emph{exclusive} one-emission ensemble.  In such an ensemble it is therefore possible for $P_{\mathrm{GI}}>P_{\mathrm{SL}}$ at the kernel level to coexist with a larger observed SL signature in a given projection: the increased GI emission probability partly migrates into vetoed higher-multiplicity configurations and is partly compensated by the enhanced Sudakov suppression in Eq.~\eqref{eq:sudakov}.  A simple toy analogue is a Poisson process with mean $\lambda$, for which the exclusive one-emission probability $\lambda e^{-\lambda}$ decreases once $\lambda$ is increased beyond unity, despite the underlying rate increasing; the shower realisation is a differential, kinematics-dependent version of this mechanism.  For this reason, Fig.~\ref{fig:qqv_srs} should be interpreted as a test of the \emph{unitarised} embedding of the longitudinal prescriptions (kernel plus Sudakov plus kinematics map), rather than as a direct point-by-point reflection of the relative size of the analytic splitting functions.

For the bosonic branchings, we perform an analogous single-resummed study in a $V+$jet topology, designed to isolate the longitudinal systematics of the $V\to V'V''$ kernels in a setting that is still sufficiently exclusive to be interpreted in terms of a single shower history.  In practice, care is required to ensure that the analysis-level reconstruction mirrors the \textsf{Herwig} shower kinematics as closely as possible and does not inadvertently bias the comparison.  First, the recoil object defining the lightlike reference direction $n^\mu$ should be identified in an infrared-safe way (e.g.\ as the hardest coloured final-state parton or via parton-level jet clustering), since overly restrictive definitions such as selecting only a gluon recoil can remove a large fraction of events and distort the reconstructed $z$ and $p_\perp$ spectra.  Second, the final-state topology used to tag the splitting channel must be made exclusive (e.g.\ exactly one $W$ and one $Z$ for $W\to WZ$, or exactly one $W^+$ and one $W^-$ for $Z/\gamma\to W^+W^-$) to avoid contamination from additional electroweak radiation, and, in particular, $Z\to W^+W^-$ and $\gamma\to W^+W^-$ cannot be separated from the final state alone and must be distinguished at the level of the generated hard process.  Finally, the kinematic variables entering the comparison should be reconstructed.

\begin{figure}[t]
\centering
\includegraphics[width=.49\textwidth]{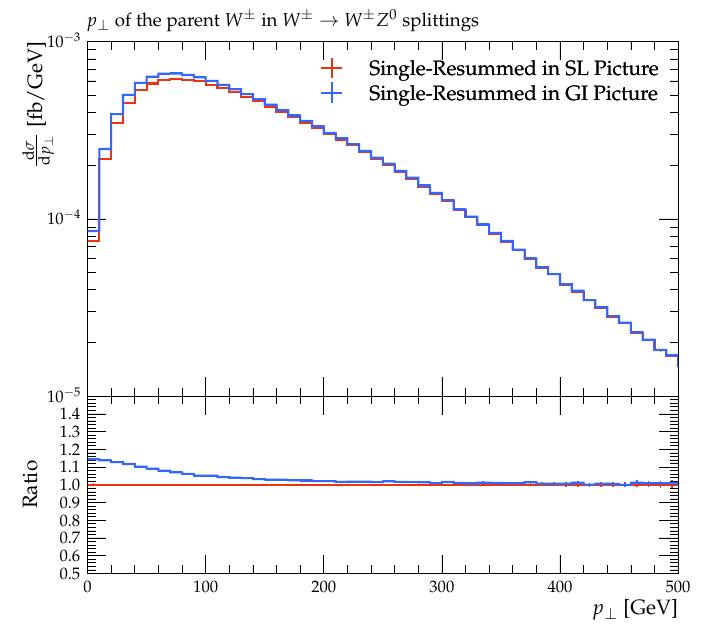}
\includegraphics[width=.49\textwidth]{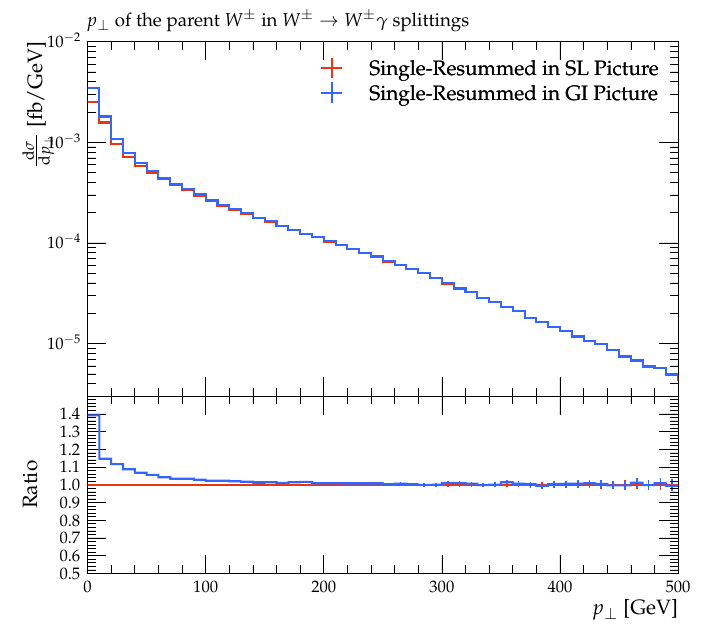}
\includegraphics[width=.49\textwidth]{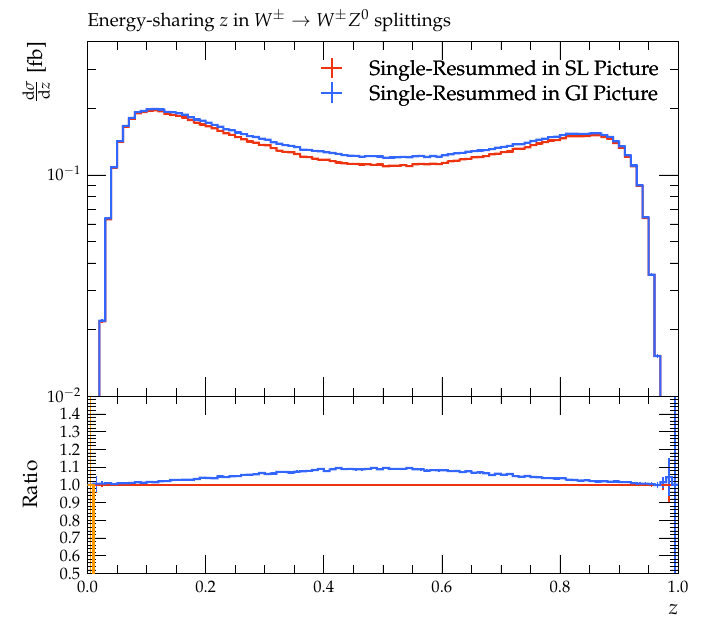}
\includegraphics[width=.49\textwidth]{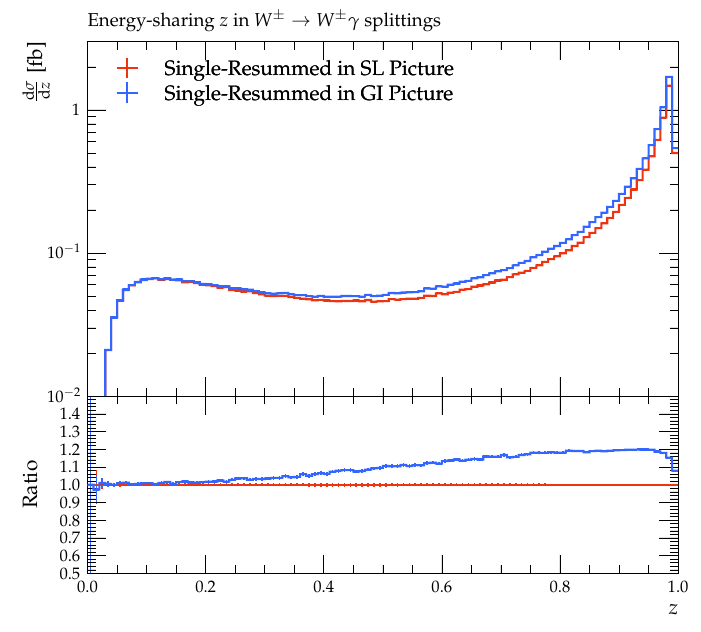}
\caption{\small SRS shower comparison between the SL and GI longitudinal prescriptions for boson-line splittings in a $pp$ sample at $\sqrt{s}=13~{\rm TeV}$. Left column: splitting-tagged $W^\pm\to W^\pm Z^0$; right column: splitting-tagged $W^\pm\to W^\pm\gamma$. Upper row: transverse-momentum spectrum of the parent $W^\pm$, $p_{\perp,W}$; lower row: reconstructed lightcone momentum fraction $z$ of the splitting as used in the \textsf{Herwig} kinematics map.}
\label{fig:vvv_srs_wj}
\end{figure}

\begin{figure}[t]
\centering
\includegraphics[width=.49\textwidth]{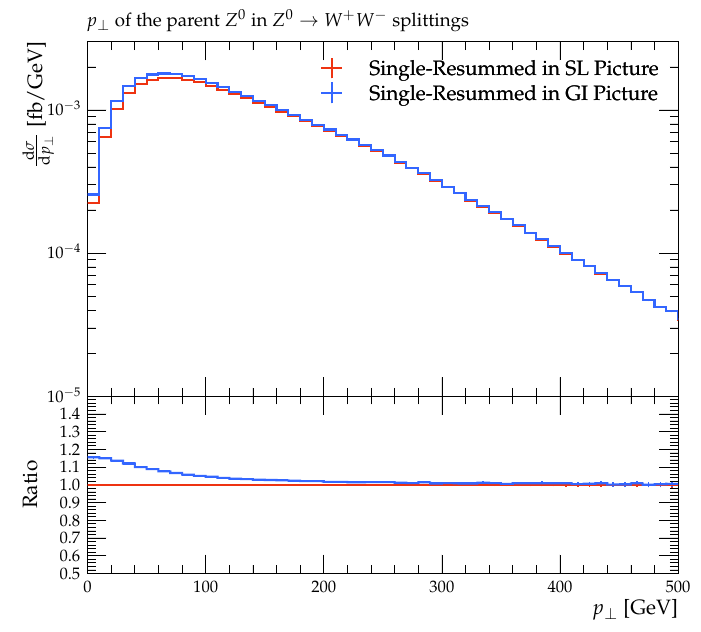}
\includegraphics[width=.49\textwidth]{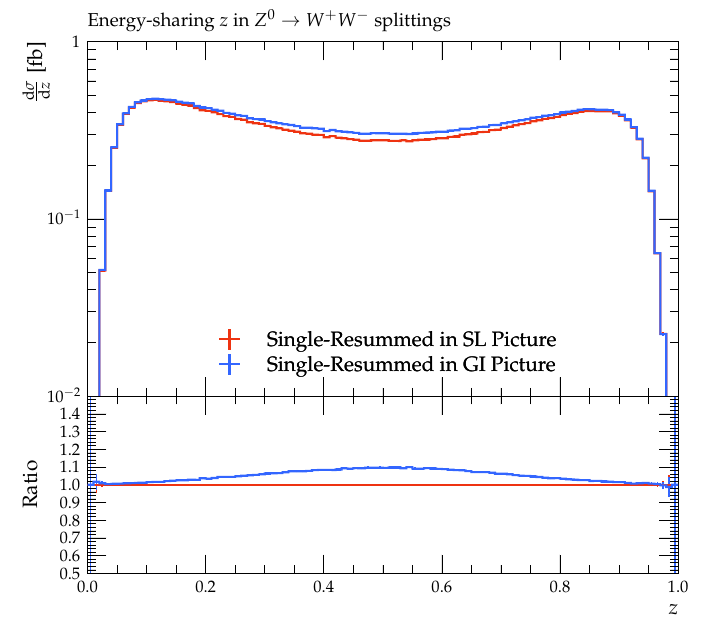}
\caption{\small As Fig.~\ref{fig:vvv_srs_wj}, but for splitting-tagged $Z^0\to W^+W^-$ branchings at $\sqrt{s}=13~{\rm TeV}$. Left: parent-$Z^0$ transverse-momentum spectrum; right: reconstructed splitting variable $z$.}
\label{fig:vvv_srs_zj}
\end{figure}

\begin{figure}[t]
\centering
\includegraphics[width=.49\textwidth]{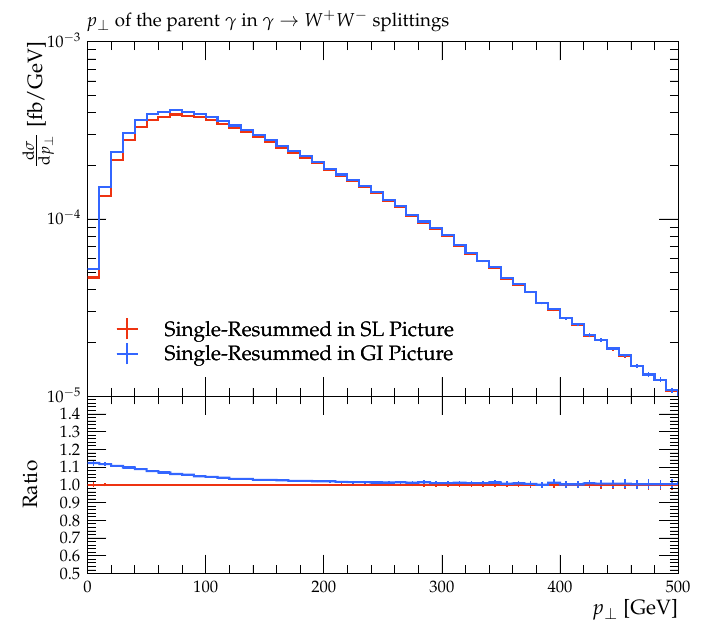}
\includegraphics[width=.49\textwidth]{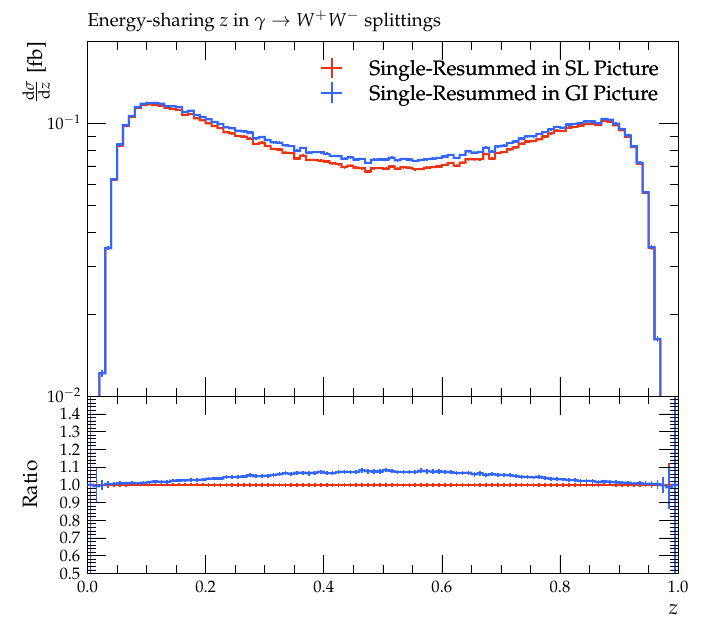}
\caption{\small As Fig.~\ref{fig:vvv_srs_wj}, but for splitting-tagged $\gamma\to W^+W^-$ branchings at $\sqrt{s}=13~{\rm TeV}$. Left: parent-$\gamma$ transverse-momentum spectrum; right: reconstructed splitting variable $z$.}
\label{fig:vvv_srs_gamj}
\end{figure}

Figures~\ref{fig:vvv_srs_wj}-\ref{fig:vvv_srs_gamj} show that, for boson-line branchings, the GI prescription yields a systematically larger single-resummed splitting-tagged rate than the SL construction, with a mild enhancement in both the parent-$p_\perp$ spectra and the reconstructed energy-sharing variable $z$ (typically at the $\mathcal{O}(5\text{-}15\%)$ level, with the largest deviations at low $p_\perp$ and away from the kinematic end-points). The effect is qualitatively similar across the channels $W^\pm\!\to W^\pm Z^0$, $W^\pm\!\to W^\pm\gamma$, $Z^0\!\to W^+W^-$ and $\gamma\!\to W^+W^-$, consistent with the fact that the comparison is dominated by the longitudinally sensitive part of the bosonic $V\!\to V'V''$ kernels rather than by the hard-process normalisation. The sign relative to the $q\!\to qV$ case is not paradoxical: in the fermion-line study, the analysis selects an exclusive one-boson ensemble, so a locally larger GI rate increases the Sudakov suppression and migrates probability into vetoed higher-multiplicity configurations, which can make the observed GI spectrum smaller even when the underlying kernel is larger. For the boson-line channels in Figs.~\ref{fig:vvv_srs_wj}-\ref{fig:vvv_srs_gamj}, the tagged $V\!\to V'V''$ branchings contribute less to the Sudakov exponent controlling the acceptance, so the change in overall no-emission probability is comparatively small and the enhanced GI longitudinal contribution (fixed by the Ward-identity/Goldstone completion) manifests directly as an upward shift in the splitting-tagged distributions, hence ${\rm GI}/{\rm SL}>1$ over most of the measured phase space.

\subsection{Full-Resummed Evolution}
\label{sec:frs}

The single-emission comparisons in Sec.~\ref{sec:srs} are designed to isolate the impact of a given helicity-dependent branching kernel in a controlled environment, but they do not, by themselves, guarantee that a modified longitudinal prescription remains well behaved once embedded into the full Markov chain. In a realistic hadron-collider evolution, longitudinal-sensitive differences propagate not only through the local form of the splitting functions but also through their interplay with QCD radiation and recoil, phase-space mapping and physicality constraints, ordering, and the cumulative Sudakov suppression from multiple emissions. These effects can either dilute kernel-level differences (via kinematic squeezing and migration in reconstructed shower variables) or amplify them (via secondary electroweak branchings and correlated changes in veto patterns). A dedicated LHC-like physics check is therefore required to validate numerical stability and to quantify how a longitudinal completion feeds into event-level systematics in the phase-space region where longitudinal effects are expected to be most visible.

Concretely, we generate $pp$ events at $\sqrt{s}=13~\mathrm{TeV}$ from a QCD hard process, $gg\to q\bar q$, using \texttt{MEQCD2to2} with \texttt{MaximumFlavour} set to 5 and imposing a minimum partonic invariant mass for the hard subprocess, $\hat M\equiv\sqrt{\hat s}>1~\mathrm{TeV}$, together with a jet hardness requirement $k_T^{\rm jets}>500~\mathrm{GeV}$. These cuts serve two purposes: first, they enforce a boosted, well-separated two-jet topology that fixes a hard starting scale for the subsequent evolution and stabilises the recoil/phase-space mapping against soft-event artefacts; second, they maximise sensitivity to longitudinal effects by pushing the shower into a regime where electroweak emissions are controlled by large logarithms, $\ln(Q^2/m_W^2)$, and where ultracollinear bosonic branchings are kinematically accessible.

We then perform six runs to disentangle intrinsic longitudinal-scheme effects from their coupling to the rest of the event generation. For each longitudinal prescription (SL and GI) we generate: (i) a reference sample with the full shower enabled (QCD$\oplus$QED$\oplus$EW) and hadronisation disabled; (ii) an evolution with EW radiation only and hadronisation disabled, which suppresses QCD-induced recoil and phase-space competition and thereby isolates the electroweak Sudakov dynamics; and (iii) a fully realistic sample with the full shower enabled (QCD$\oplus$QED$\oplus$EW) and hadronisation enabled, which tests the persistence of the observed differences once non-perturbative modelling and hadronic activity are reinstated. In all cases, electroweak radiation is enabled on both fermion and boson lines via the $q\to q'W/Z$ and $V\to V'V''$ Sudakov modules, and the SL versus GI comparison is obtained by switching the corresponding \texttt{LongitudinalEWScheme} settings while keeping all other generator parameters fixed.

The resulting samples are analysed with dedicated \textsf{Rivet} routines which identify $W/Z$ occurrences in the event record and construct inclusive spectra and reconstructed shower variables (including the energy-sharing variable $z$) for representative bosonic channels. In the full-resummed context, this provides an internal consistency check of the longitudinal completion: observables dominated by last-copy, on-shell gauge bosons mainly probe the choice of physical polarisation basis and are therefore expected to agree between SL and GI up to subleading recoil and bookkeeping effects. By contrast, splitting-tagged configurations probe iterated, generally off-shell gauge-boson currents inside the shower, whose longitudinal component is fixed by the broken-theory Ward identity and its associated Goldstone-matching; it is therefore in such samples, particularly in bosonic channels and ultracollinear regions, that the largest sensitivity to the GI completion is expected.

\begin{figure}[t]
\centering
\includegraphics[width=.49\textwidth]{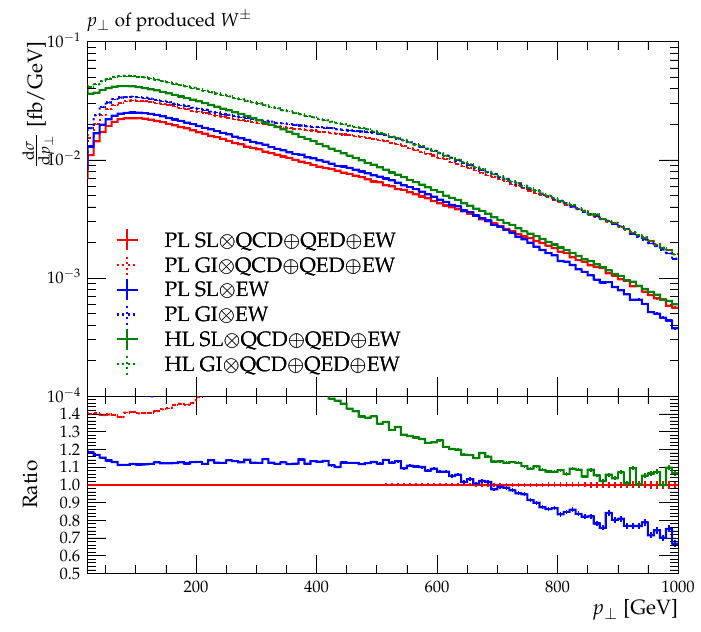}
\includegraphics[width=.49\textwidth]{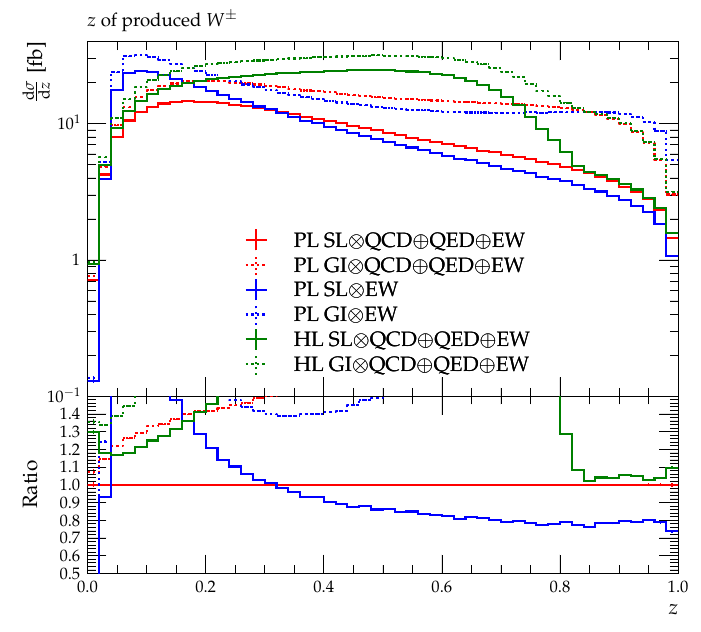}
\includegraphics[width=.49\textwidth]{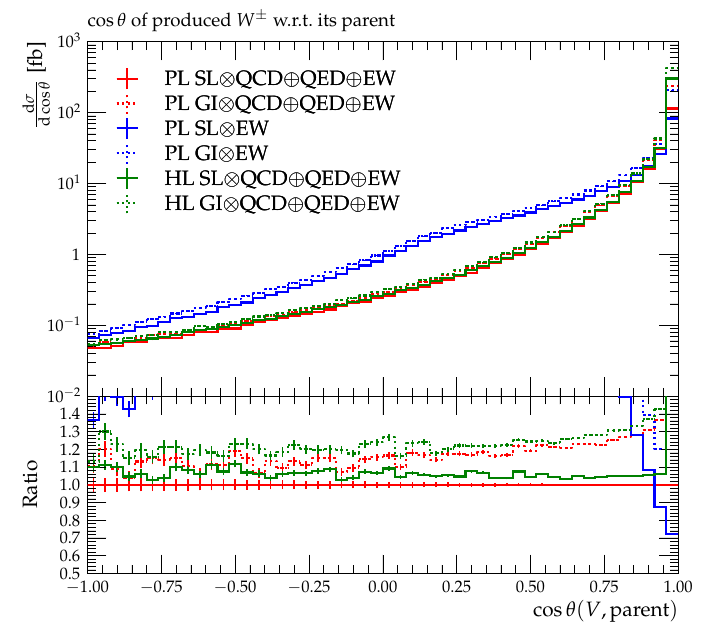}
\includegraphics[width=.49\textwidth]{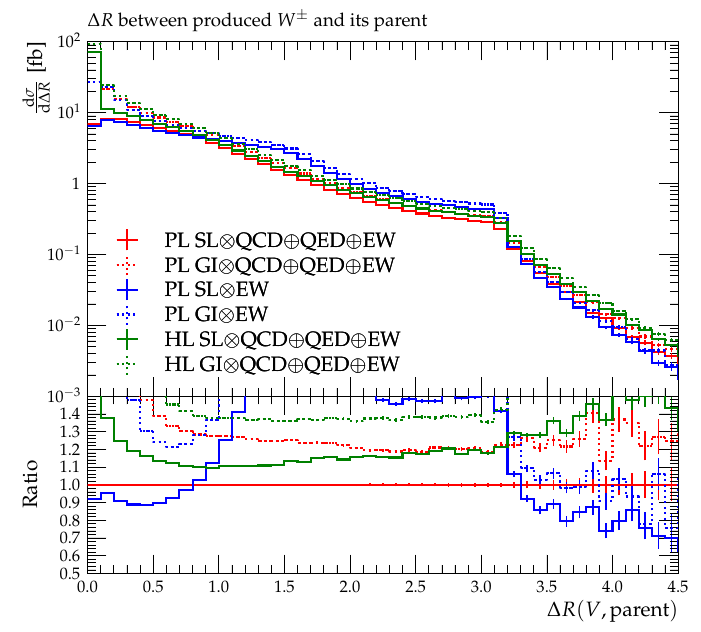}
\caption{\small Full-resummed evolution at $\sqrt{s}=13~\mathrm{TeV}$: kinematics of produced $W^\pm$ in a pure QCD dijet hard-process configuration ($gg\to q\bar q$ with \texttt{MaximumFlavour = 5}, $\hat M>1~\mathrm{TeV}$ and $k_T^{\rm jets}>500~\mathrm{GeV}$). Shown are, from left to right and top to bottom, the $p_\perp$ spectrum, the reconstructed energy-sharing variable $z$, $\cos\theta(V,\mathrm{parent})$, and $\Delta R(V,\mathrm{parent})$, where the ``parent'' is the first non-self ancestor in the event record. Curves compare the SL and GI longitudinal prescriptions for parton-level (PL, hadronisation off) and hadron-level (HL, hadronisation on) runs, using either the full shower (QCD$\oplus$QED$\oplus$EW) or EW radiation only.}
\label{fig:rs_w}
\end{figure}

\begin{figure}[t]
\centering
\includegraphics[width=.49\textwidth]{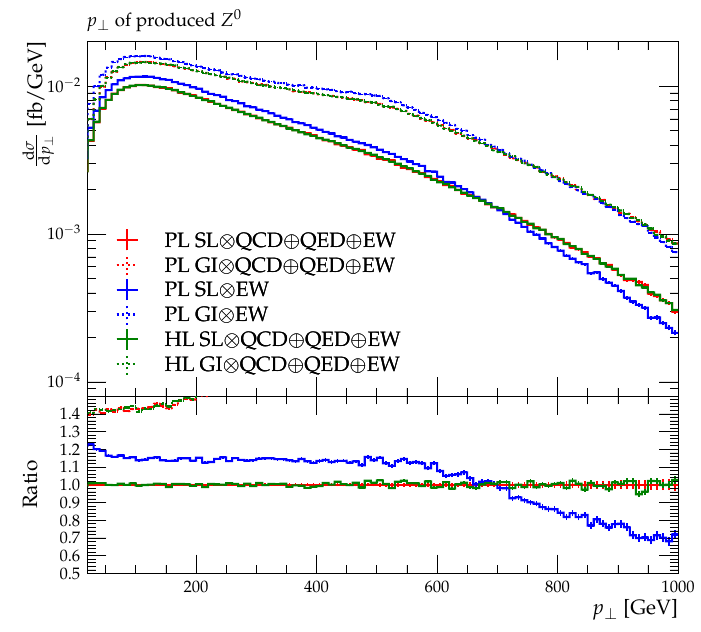}
\includegraphics[width=.49\textwidth]{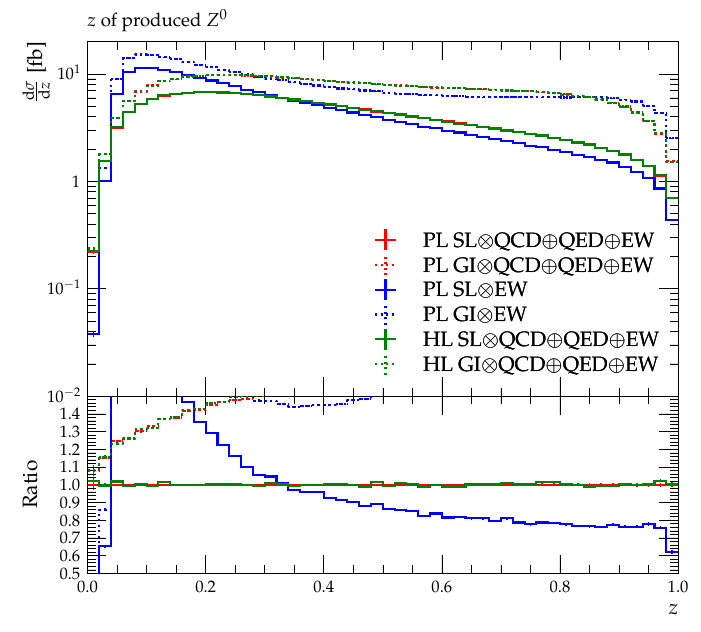}
\includegraphics[width=.49\textwidth]{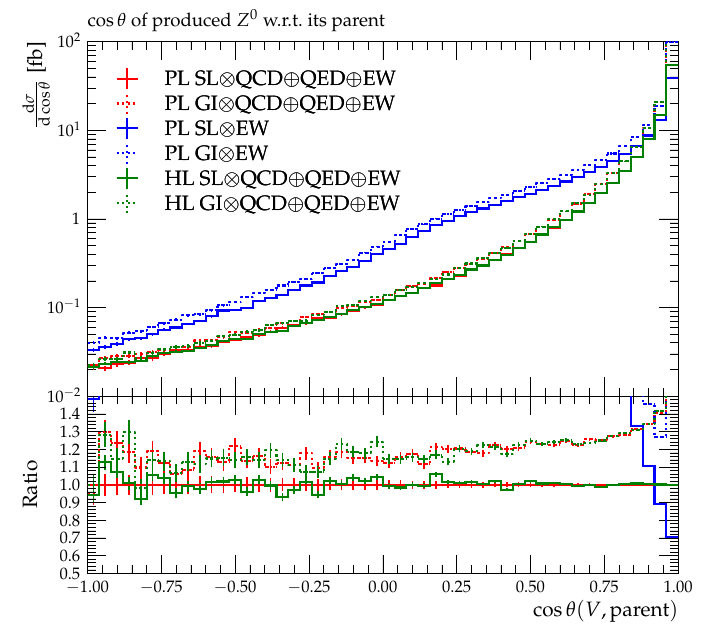}
\includegraphics[width=.49\textwidth]{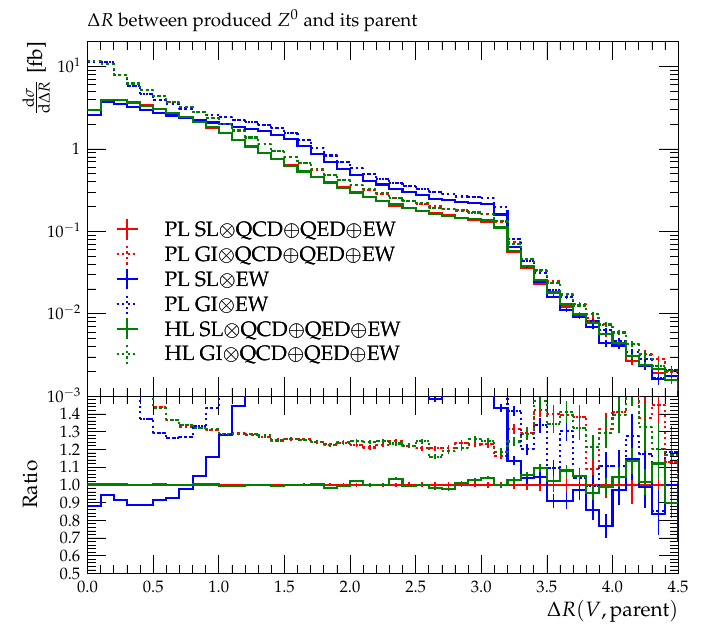}
\caption{\small As Fig.~\ref{fig:rs_w}, but for produced $Z^0$.}
\label{fig:rs_z}
\end{figure}

Figures~\ref{fig:rs_w}-\ref{fig:rs_z} demonstrate that the SL-GI differences observed at single emission persist in the full Markov evolution as controlled, smooth modifications. The angular observables $\cos\theta(V,\mathrm{parent})$ and $\Delta R(V,\mathrm{parent})$ remain essentially unchanged under SL$\leftrightarrow$GI, while $p_\perp$ and the reconstructed energy-sharing variable $z$ show moderate rate/tilt-like shifts whose size depends on whether the full shower (QCD$\oplus$QED$\oplus$EW) or EW radiation only is used. In particular, the near-indistinguishability of the $\Delta R$ and $\cos\theta$ shapes under the SL $\leftrightarrow$ GI alteration implies that the radiation pattern remains coherent: the angular structure encoded by the shower recoil map and ordering is not destabilised by the longitudinal prescription, and any scheme dependence is largely absorbed into an overall rate/weight redistribution rather than an angular reconfiguration.

Turning to the differences between the full shower (QCD$\oplus$QED$\oplus$EW) and EW radiation only, and between parton- and hadron-level runs, the most pronounced effect is the large migration in the $z_W$ spectrum when hadronisation is enabled, while the corresponding $p_{\perp,W}$ and angular observables remain comparatively stable. This behaviour is expected because the plotted $z$ is a \emph{reconstructed} lightcone fraction defined with respect to the first non-self ancestor in the event record; once hadronisation is switched on, the emitting coloured system undergoes non-perturbative cluster formation and momentum reshuffling, so that the record-level ``parent'' providing the reference direction and lightcone normalisation is no longer a faithful proxy for the pre-hadronisation shower emitter, leading to a substantial distortion of the inferred $z$ even when the boson four-momentum itself is only weakly affected. For the $Z^0$ sample, by contrast, the spectra exhibit negligible hadronisation sensitivity, consistent with the fact that the dominant $Z$ population is generated at comparatively hard, perturbative stages of the evolution where the ancestry assignment remains anchored to a well-defined partonic emitter; consequently, the record-level reconstruction of $z_Z$ is far less exposed to non-perturbative recoil and bookkeeping effects.

\section{Conclusions}
\label{sec:conc}

We have constructed and implemented a GI completion for EW parton-shower branchings in the broken theory, formulated at the level of helicity-dependent quasi-collinear splitting operators. The construction enforces the Ward-identity constraints on longitudinal currents by explicit Goldstone-matching, thereby removing the scheme ambiguity associated with the naive $p^\mu/m_V$ growth of $\varepsilon^\mu_L$ while retaining a probabilistic shower interpretation and spin correlations. We provided the reduced helicity amplitudes for $q\to q'V$ and $V\to V'V''$ in the \textsf{Herwig~7} conventions and exposed the minimal modification pattern: only the longitudinal $0_*$ component receives GI contributions, controlled by the Goldstone coefficient and Yukawa structures, leaving the transverse sector unchanged. In practical terms, this means inclusive observables dominated by promptly contracted, near on-shell vector-boson matrix elements are largely unaffected, while splitting-tagged and reconstructed shower-current observables provide the cleanest handles on the scheme choice. The corresponding implementation is numerically stable and configurable through run-card switches, enabling systematic comparisons with the widely used SL prescription and providing a convenient starting point for extensions to non-standard chiral and triple-gauge couplings.

At the phenomenological level, we validated the impact of the scheme choice in both single-emission and full-resummed evolutions. The single-resummed studies isolate kernel-level effects and clarify how changes in the longitudinal operator can migrate between exclusive and inclusive observables through Sudakov suppression. In the fully iterated shower evolution, we find that switching SL $\leftrightarrow$ GI produces consistent, controlled shifts without disrupting the coherence of the radiation pattern: $\Delta R$ and $\cos\theta$ distributions remain stable, indicating that the recoil map, ordering and physicality constraints are not destabilised by the completion. Differences between shower environments (full QCD$\oplus$QED$\oplus$EW versus EW radiation only) primarily reflect the expected phase-space competition and recoil coupling to QCD activity. Finally, we observe that hadronisation can strongly affect reconstructed shower variables that reference an event-record parent (notably $z_W$), while leaving harder, perturbatively anchored spectra (e.g.\ the $Z^0$ observables considered here) largely unchanged. Overall, the GI completion provides a theoretically consistent and practically robust baseline for longitudinal EW radiation in general-purpose event generation, reducing unphysical scheme artefacts precisely in the ultracollinear regime where such effects are parametrically enhanced. A natural next step is to extend the validation to a broader set of multi-emission LHC analyses (including decays and matching choices) in order to map where longitudinal systematics propagate into experimentally used observables.

We emphasise that the SL-GI difference is parametrically controlled by $m_{1,2}^2/\tilde q^{\,2}$, so the two prescriptions must coincide at high evolution scales while deviations are expected in the low-$\tilde q$ region where symmetry-breaking terms become numerically relevant. In that region, the shower cutoff/physicality constraints and the exclusivity of the one-emission selection can amplify small kernel-level shifts into non-monotonic changes in accepted event samples through Sudakov suppression and vetoed higher-multiplicity configurations.

\section*{Acknowledgements}
\noindent We thank our fellow \textsf{Herwig~7} authors for useful discussions. This work has received funding from the European Union's Horizon 2020 research and innovation programme as part of the Marie Skłodowska-Curie Innovative Training Network MCnetITN3 (grant agreement no.~722104). \textit{MRM} is also supported by the UK Science and Technology Facilities Council (grant numbers ST/T001011/1 and ST/X000745/1). 

\appendix
\section{Run-card interface for the Longitudinal EW Schemes}
\label{sec:ui-long}

The GI longitudinal completion can be enabled independently for fermion emissions $q\to q'V$ and bosonic splittings $V\to V'V''$ via the following run-card switches. The longitudinal prescription is selected through
\begin{verbatim}
set /Herwig/Shower/<Splitting class>:LongitudinalEWScheme <picture>
\end{verbatim}
with relevant classes being \texttt{QtoQWZSudakov} for $q\to q'V$ splittings, \texttt{VtoVVSudakov} for $V\to V'V''$ splittings, and \texttt{GammatoWWSudakov} for $\gamma\to W^+W^-$ splitting. The \texttt{<picture>} can be set to either \texttt{Subtraction} or \texttt{GaugeInvariant}. 
In the GI case, the relative coefficient $c_G$ multiplying the Goldstone-matching contribution can be configured via
\begin{verbatim}
set /Herwig/Shower/<Splitting class>:GI.cG <value>
\end{verbatim}
with the default set to $c_G=1$. Setting $c_G=0$ switches off the Goldstone-matching term, leaving only the subtraction remainder in the longitudinal current. Other values ($0<c_G<1$, $c_G>1$, or $c_G<0$) may be used as a purely diagnostic interpolation/stress test, but they do not correspond to a gauge-invariant SM prescription. In practice, the spread between $c_G=1$ and $c_G=0$ can be taken as an estimate of the residual scheme/model dependence associated with the treatment of longitudinal emissions: this difference is controlled by symmetry-breaking mass effects and therefore decreases with increasing evolution scale.

By default the Goldstone Yukawa couplings are derived internally from the SM masses and EW inputs; they can be overridden (for BSM studies or cross-checks) through
\begin{verbatim}
set /Herwig/Shower/QtoQWZSudakov:GBYukawa.Left.Re  <value>
set /Herwig/Shower/QtoQWZSudakov:GBYukawa.Left.Im  <value>
set /Herwig/Shower/QtoQWZSudakov:GBYukawa.Right.Re <value>
set /Herwig/Shower/QtoQWZSudakov:GBYukawa.Right.Im <value>
\end{verbatim}
Keeping all four values at zero restores the default SM-derived Yukawas. For BSM chiral vector couplings at the $qqV$ vertex, one may instead import numerical left/right couplings through
\begin{verbatim}
set /Herwig/Shower/QtoQWZSudakov:CouplingValue.Left.Re  <value>
set /Herwig/Shower/QtoQWZSudakov:CouplingValue.Left.Im  <value>
set /Herwig/Shower/QtoQWZSudakov:CouplingValue.Right.Re <value>
set /Herwig/Shower/QtoQWZSudakov:CouplingValue.Right.Im <value>
\end{verbatim}
in which case, the internal SM values are bypassed for the affected channels. For BSM studies of the triple-gauge coupling normalisation, a numerical complex coupling can be provided via
\begin{verbatim}
set /Herwig/Shower/VtoVVSudakov:CouplingValue.Re <value>
set /Herwig/Shower/VtoVVSudakov:CouplingValue.Im <value>
\end{verbatim}
Again, keeping both at zero restores the default SM couplings.

% References 
\bibliography{references}

\end{document}